# Delocalized Photonic Deep Learning on the Internet's Edge

Alexander Sludds[1], Saumil Bandyopadhyay[1], Zaijun Chen[1], Zhizhen Zhong[2], Jared Cochrane[1,3], Liane Bernstein[1], Darius Bunandar[1], P. Ben Dixon[3], Scott A. Hamilton[3], Matthew Streshinsky[4], Ari Novack[4], Tom Baehr-Jones[4], Michael Hochberg[4], Manya Ghobadi[2], Ryan Hamerly[1,5], and Dirk Englund[1]

[1]*Research Laboratory of Electronics, MIT, Cambridge, MA, 02139, USA*
[2]*Computer Science and Artificial Intelligence Laboratory, MIT, Cambridge, MA, 02139, USA*
[3]*Lincoln Laboratory, MIT, Lexington, MA, 02421, USA*
[4]*Nokia Corporation, 171 Madison Avenue Ste 1100, New York, NY, 10016, USA and*
[5]*NTT Research Inc., PHI Laboratories, 940 Stewart Drive, Sunnyvale, CA 94085, USA*

Advances in deep neural networks (DNNs) are transforming science and technology [1–4]. However, the increasing computational demands of the most powerful DNNs limit deployment on low-power devices, such as smartphones and sensors – and this trend is accelerated by the simultaneous move towards Internet-of-Things (IoT) devices. Numerous efforts are underway to lower power consumption, but a fundamental bottleneck remains due to energy consumption in matrix algebra [5], even for analog approaches including neuromorphic [6], analog memory [7] and photonic meshes [8]. Here we introduce and demonstrate a new approach that sharply reduces energy required for matrix algebra by doing away with weight memory access on edge devices, enabling orders of magnitude energy and latency reduction. At the core of our approach is a new concept that decentralizes the DNN for delocalized, optically accelerated matrix algebra on edge devices. Using a silicon photonic smart transceiver, we demonstrate experimentally that this scheme, termed Netcast, dramatically reduces energy consumption. We demonstrate operation in a photon-starved environment with 40 aJ/multiply of optical energy for 98.8% accurate image recognition and <1 photon/multiply using single photon detectors. Furthermore, we show realistic deployment of our system, classifying images with 3 THz of bandwidth over 86 km of deployed optical fiber in a Boston-area fiber network. Our approach enables computing on a new generation of edge devices with speeds comparable to modern digital electronics and power consumption that is orders of magnitude lower.

On present-day computing devices, the critical bottleneck in DNN inference tasks lies in the need to evaluate large matrix algebra. This bottleneck has motivated new analog computing architectures including neuromorphic, analog memory and photonic meshes [6, 8, 9]. However, in all of these approaches, memory access and multiply-accumulate (MAC) functions remain a stubborn bottleneck near 1 pJ per MAC [5, 10–13]. Edge devices typically use chip-scale sensors, occupy millimeter-scale footprints and consume milliwatts of power. Their small footprint and low power budget mean performance is limited by the size, weight, and power (SWaP) of computing systems integrated on the device.

To make advanced DNNs at all feasible on low-power devices, industry has resorted to *offloading* compute-heavy DNN inference to cloud servers. For instance, as illustrated in Figure 1, a human device may send a voice query as a vector $U$ to a cloud server, which returns the inference result $V$ to the client. This offloading architecture adds a $\sim$ 200 ms latency to voice commands [14] which makes services such as self-driving impossible. Moreover, offloading adds security risks in both the edge and cloud: hacking of the communication of client data (in vector $U$) has led to security breaches of private data. If local computation is used to avoid this problem then models which are very expensive to train can be duplicated and stolen by competitors [15, 16].

To address these problems, we introduce here a photonic edge computing architecture, named "Netcast," to minimize the energy and latency of large linear algebra operations such as general matrix-vector multiplication (GEMV) [5]. In the Netcast architecture, cloud servers stream DNN weight data ($W$) to edge devices in an analog format for ultra-efficient optical GEMV that eliminates all local weight memory access [17].

As illustrated in Fig. 1, servers containing a 'smart transceiver' [18] – which may be in the standard pluggable transceiver format represented in Fig. 1(a) – periodically broadcast the weights ($W$) of commonly used DNNs to edge devices, using wavelength division multiplexing (WDM) to leverage the large spectrum available at the local access layer. Specifically, the $(M,N)$-sized weight matrix of one DNN layer may be encoded in a time-frequency basis by the amplitude-modulated field $W_n(t) = \sum_{j=1}^{M} w_{nj} e^{-i\omega_n t} \delta(t - j\Delta T)$, where the optical amplitude $w_{nj}$ at frequency $\omega_n$ and time step $j$ represents the $n$th row of the weight matrix illustrated in Fig. 1(d) and $\delta$ is the impulse response function.

Suppose now that a camera in Fig. 1 requires inference on an image $X$. To do so, it waits for the server to stream the 'image recognition' DNN weights, which it modulates with $X(t) = \sum_{j=1}^{M} x_j \delta(t - j\Delta T)$ using a broadband optical modulator and subsequently separates the wavelengths to $N$ time-integrating detectors to produce the vector-vector dot product $Y_n(t) = \sum_{j=1}^{M} w_{nj} x_j \delta(t - j\Delta t)$. This architecture minimizes the active components at the client, requiring only a single optical modulator, digital-to-analog converter (DAC) and analog-to-digital converter (ADC).

## EXPERIMENT

We demonstrate the Netcast protocol with a smart transceiver, shown in Fig. 2(a), made in a commercial silicon-photonic CMOS foundry (OpSIS/IME, described in Supplementary Materials II). The smart transceiver is composed of 48 Mach-Zehnder modulators (MZMs), each capable of modulation up to 50 Gbps for a total bandwidth of 2.4 Tbps [19]. The smart transceiver supports WDM, with Fig. 2(b) showing 16 WDM lasers simultaneously transmitting through the chip with $\approx -10$ dBm (100 $\mu$W) power per wavelength. Fig. 2(c) shows an open eye diagram at 50 GHz (Supplementary Materials VIII). Weights are transmitted over 86 km of deployed optical fiber, shown in Fig. 2(d), between MIT's main campus to MIT Lincoln Laboratory and back to main campus. The client, shown in Fig. 2(e), applies input activation values to the incoming weight data using a high-speed (20 GHz) broadband lithium niobate MZM, with Fig. 2(f) showing an open eye diagram at 10 GHz (limited by testing equipment). A passive wavelength demultiplexer separates each wavelength channel for detection onto an array of custom time-integrating receivers, with an example of time integration shown in Fig. 2(g)(Supplemental Materials VI). After integration, the generated voltages from the receivers are measured by a digitizer and stored in memory. Additional post-processing steps, such as the nonlinear activation function, are performed using a computer.

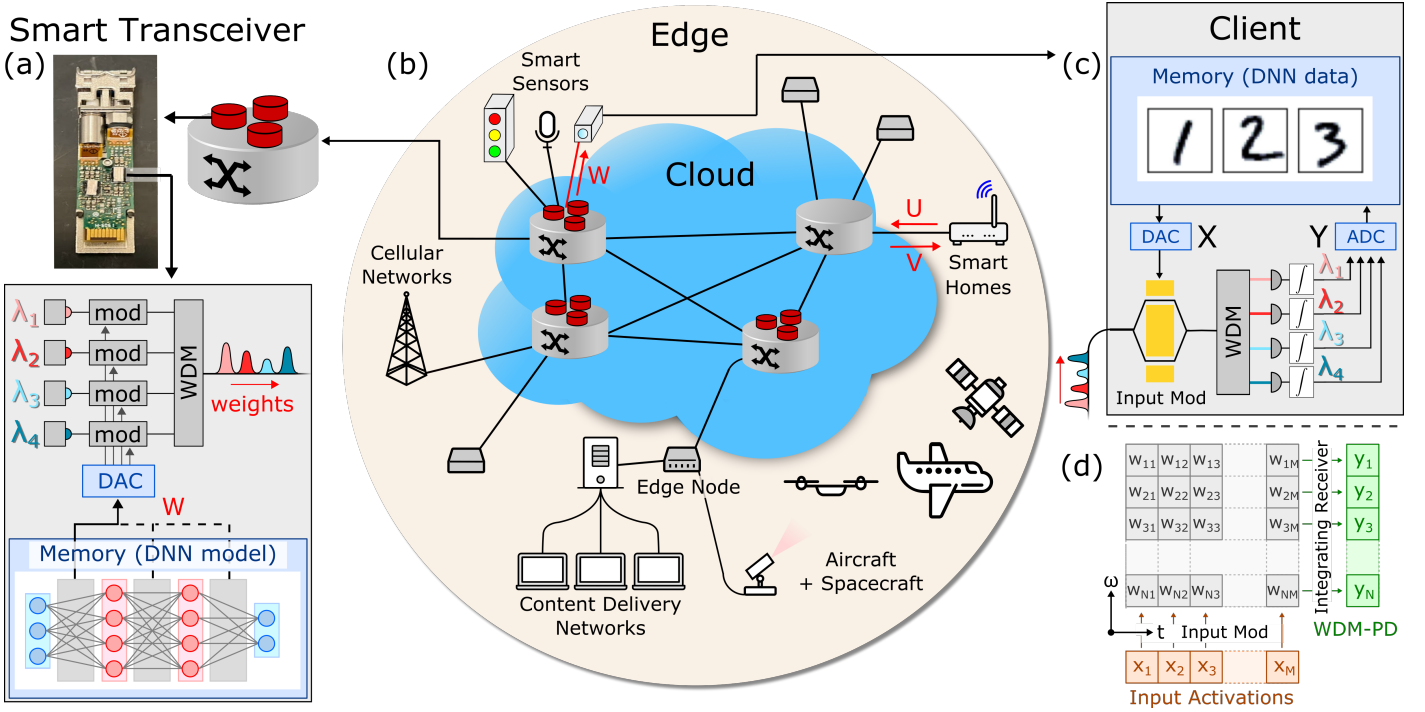

Figure 1. Netcast concept. (a) Smart transceivers integrated alongside cloud computing infrastructure including servers, data storage, network switches, and edge nodes. The smart transceiver sequentially encodes layers of a neural network model onto the intensity of distinct optical wavelengths using digital-to-analog converters (DACs), optical modulators, and lasers. Wavelength-division multiplexers (WDMs) combine the separate wavelengths from each modulator to the smart transceiver output. (b) U and V highlight current solutions to large model deployments on the edge, with edge device data communicated back to cloud computers. In our solution, smart transceivers have connections to many devices at the edge of the communications network including cellular networks, smart sensors, content delivery networks and aircraft. (c) The edge client encodes input activation data onto a single broadband optical modulator, modulating all weight wavelengths simultaneously. Wavelengths are separated with a WDM and the result of matrix-vector multiplication is computed on time-integrating receivers. (d) Matrix-vector products between an $M$-element input vector and $(M, N)$ weight matrix are time(t)-frequency($\omega$) encoded, with each wavelength accumulating its results on a time-integrating receiver.

In Fig. 3(a), we show the flow of data through the experimental setup and the accuracy it can achieve. Weight data are encoded to multiple modulators simultaneously. For clarity, we show a single row of the digit "3" being encoded and the resulting time trace from a single wavelength. We demonstrate computing with high accuracy, with Fig. 3(b) showing 8 bits of precision, more than the $\approx$5 bits of precision required for neural network computation [20, 21]. After calibrating the system we perform image classification by running a benchmark handwritten digit classification task (MNIST) which was trained on a digital computer (Supplementary Materials XVI, XIV). Fig. 3(c) illustrates an example of the systems computing result for classifying the digit "3". We then test the system's performance both locally and over deployed fiber using a benchmark 3-layer MNIST model with 100 neurons per hidden layer (Supplementary Materials XIV). Using 1,000 test images locally, we demonstrate 98.7% accurate computation, comparable with the model's baseline accuracy of 98.7%. Using the same test images, we utilize 3 THz of bandwidth over the deployed fiber and classify MNIST digits with 98.8% accuracy. This result shows the potential for this architecture to support ultra-high bandwidths in real-world deployed systems using conventional components.

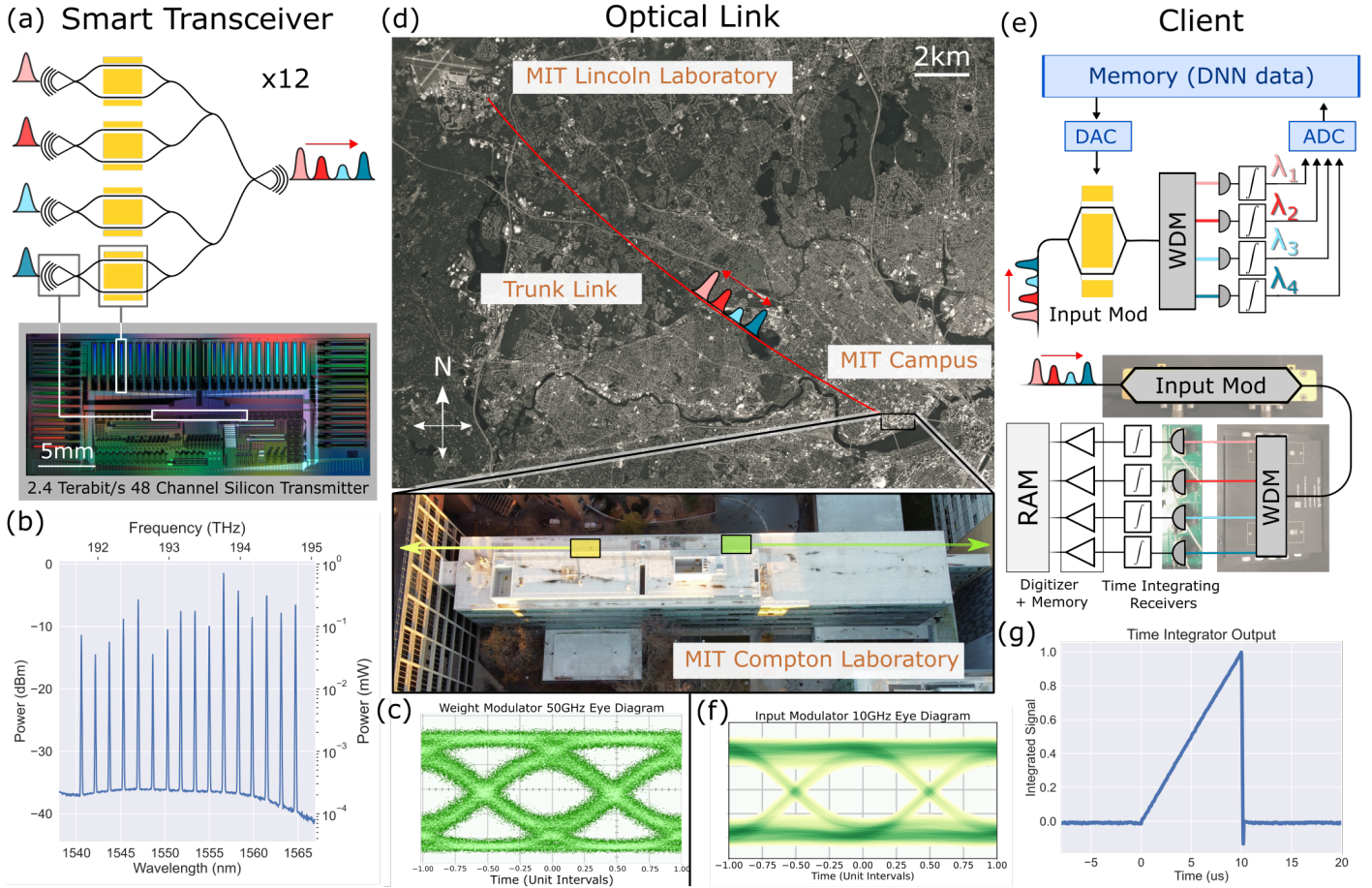

Figure 2. Experimental demonstration of Netcast system. (a) Smart transceiver comprised of a 48-modulator silicon photonic transmitter with 2.4 Tbps of total bandwidth. (b) Optical spectrum of smart transceiver output, showing 16 laser sources across 3 THz of bandwidth with >25 dB optical SNR. (c) An example of high-speed operation of the smart transceiver modulators, with a 50 GHz open eye. (d) Weights are sent over 86 km of deployed optical fiber connecting the smart transceiver to the client. (e) Client receiver composed of a broadband, high-speed optical modulator, a WDM demultiplexer, and custom time-integrating receivers. (f) The client input modulator also achieves an open eye of 10 GHz (test equipment limited). (g) Example time-integrating receiver waveform showing constant optical power being accumulated over 10 $\mu s$ and resetting. Satellite imagery in (d) taken using deployed satellite (Planet.com).

## ENERGY EFFICIENCY

Netcast is designed to minimize the power used at the client. To enable this, we make sure every component at the client is performing a large number of MACs (M or N) for modulation and electrical readout respectively. Only a single MZM and DAC are used to encode input data across $N$ wavelengths, enabling $N$ MACs of work for every voltage applied to the modulator. While the energy costs of these individual components can be high, they have high parallelism, performing many MACs of work per time step. For encoding input activations, the client only uses a single broadband optical modulator, allowing for $\approx (1/N)$ pJ/MAC of energy consumption using standard components. Furthermore, the integrator and ADC can be much slower than the speed of modulated weights, since readout occurs after $M$ timesteps. As a result, the integrator and ADC can be $M$ times slower, decreasing the cost of electrical readout components to $\approx (1/M)$ pJ/MAC. Assuming near-term values of $N = M = 100$, client energy consumption can reach $\approx$10 fJ/MAC, which is three orders of magnitude lower than possible in existing digital CMOS. The scaling of the client energy consumption is summarized in Table I.

| Netcast Client Energy Consumption | | | | |
|---|---|---|---|---|
| Device | Number of Devices | Fan-out | Energy per Device | Energy per MAC |
| Modulator [19] | 1 | $N$ | $\sim 1$ pJ | $\sim (1/N)$ pJ |
| DAC [22] | 1 | $N$ | $\sim 1$ pJ | $\sim (1/N)$ pJ |
| ADC [23] | 1 | $M$ | $\sim 1$ pJ | $\sim (1/M)$ pJ |
| Integrator [24] | $N$ | $M$ | $\sim 1$ fJ | $\sim (1/M)$ fJ |
| Total | – | – | – | $\sim (1/N)$ pJ |

Table I. Device contributions to receiver performance assuming conventional technology. Device energy consumption is amortized by either a spatial fan-out factor ($N$) or time-domain fan-out factor ($M$). We assume a carrier depletion modulator in silicon is used and that a single high-speed (GHz) ADC reads out from an array of $N$ slow integrators.

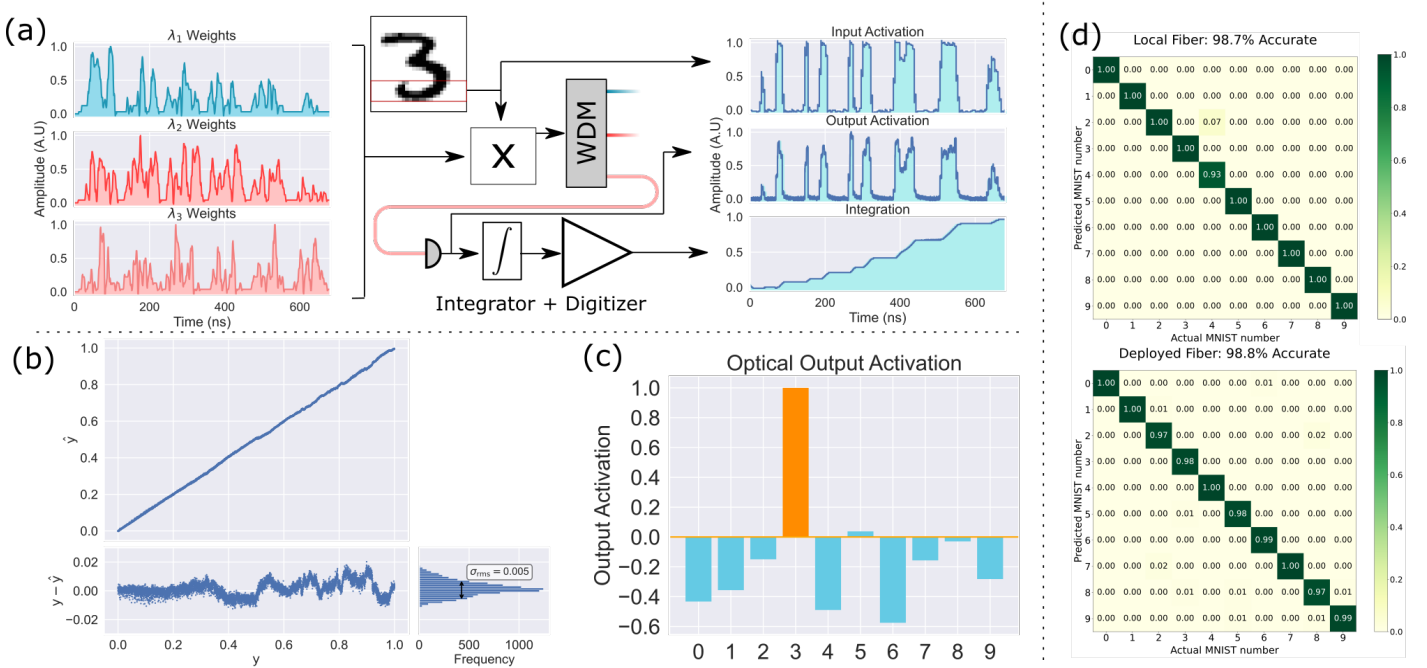

Figure 3. Computational accuracy of Netcast system. (a) Weight data from multiple wavelength channels is simultaneously modulated by input data. After wavelength multiplexing the generated photocurrent is time-integrated. (b) Floating-point computing accuracy comparing the results of 10,000 scalar-scalar floating point multiplications. Electrical floating point results are designated as $y$ and optical results are designated as $\hat{y}$. The difference $y - \hat{y}$ has a standard deviation of $\sigma_{rms} = 0.005$ or $\approx$ 8-bit accuracy. (c) Example output activation data from the optical setup correctly classifying the digit 3. (d) Computing results of image classification over both local links and the 86 km deployed fiber link.

## RECEIVER SENSITIVITY

One limitation of the performance of all optoelectronic neural networks is the finite signal to noise ratio (SNR) of modern photoreceivers [25]. Loss from fiber propagation or diffraction in free space links can force the client to operate in a photon starved environment. To enable the best use of the received optical signal we must have the lowest possible noise floor, ideally operating at a shot-noise limit with $\approx$ 1 photon / MAC. Modern photoreceivers are limited by either thermal noise of readout electronics (also called Johnson-Nyquist noise [26]), shot noise, flicker ($1/f$) noise, and relative intensity (RIN) noise of the laser; of these, for integrated optoelectronics, thermal and shot noise are dominant in Netcast (see Supplementary Material XIII,XXI). We overcome this problem with time-integrating receivers, which accumulate partial results from vector-matrix multiplication. In Fig. 4(a) we compare the sensitivity of different photoreceivers. Amplified photoreceivers, shown on the right of Fig. 4(a), have typical sensitivities of $\approx$10-100 fJ/MAC. Amplified linear mode avalanche photodetectors, shown in the middle of Fig. 4(a), overcome some of the thermal noise of the amplifier and achieve $\approx$1fJ/MAC. Our custom time-integrating receivers, shown on the left of Fig. 4(a), boost the measured signal by $M$ at readout, increasing the voltage SNR by $M$ and enabling receivers operating with $\approx$10 aJ/MAC ($\approx$100 photons) of optical energy. This result brings Netcast close to the fundamental quantum limit of optical computation [27, 28], which we can reach by engineering the receiver to lower thermal noise.

Thermal noise is a hardware dependent noise source, originating from the thermal motion of charge carriers in an electrical conductor. In an RC circuit, thermal noise manifests in a fluctuation in the number of readout electrons in a circuit given by $\sigma_{th} = \sqrt{kTC}/q$ where $k$ is Boltzmann's constant, $T$ is temperature, $q$ is the electron charge and $C$ is the capacitance of the receiver [29]. Conventional amplified photodetectors used for neural network computation must exceed a required SNR every time a MAC is performed. Time-integrating receivers, however, only need to meet a similar SNR after performing $M$ MACs. As a result, time-integrating receivers will generate $E_{mac}\eta\frac{M}{h\nu}$ electrons after $M$ timesteps, where $\eta$ is the quantum efficiency of the detector and $h\nu$ is the photon energy. This leads to an SNR from a time-integrator of $\frac{E_{mac}\eta\frac{M}{h\nu}M}{\sqrt{kTC}}$, which is $M$ times higher than the SNR from a single MAC of an amplified receiver.

Improvements to time integrating receivers are possible by minimizing the integration capacitance of the receiver. Fig. 5(a) shows the thermal noise limit of time integrating receivers as integration capacitance is decreased. This noise floor is fundamentally connected to the size-scale of photodetectors, readout electronics and their proximity of integration [11]. Modern foundry processes enable $\approx$ 1fF scale receivers, lowering the thermal readout noise to the single photon per MAC level [30–32]. This single photon per MAC regime is fundamentally limited by the quantum nature of light, where precision is determined by the Poissonian distribution of photons that arrive within a measurement window. Poissonian noise, also called shot noise, can be observed in experimentally measured data in Fig. 5(c).

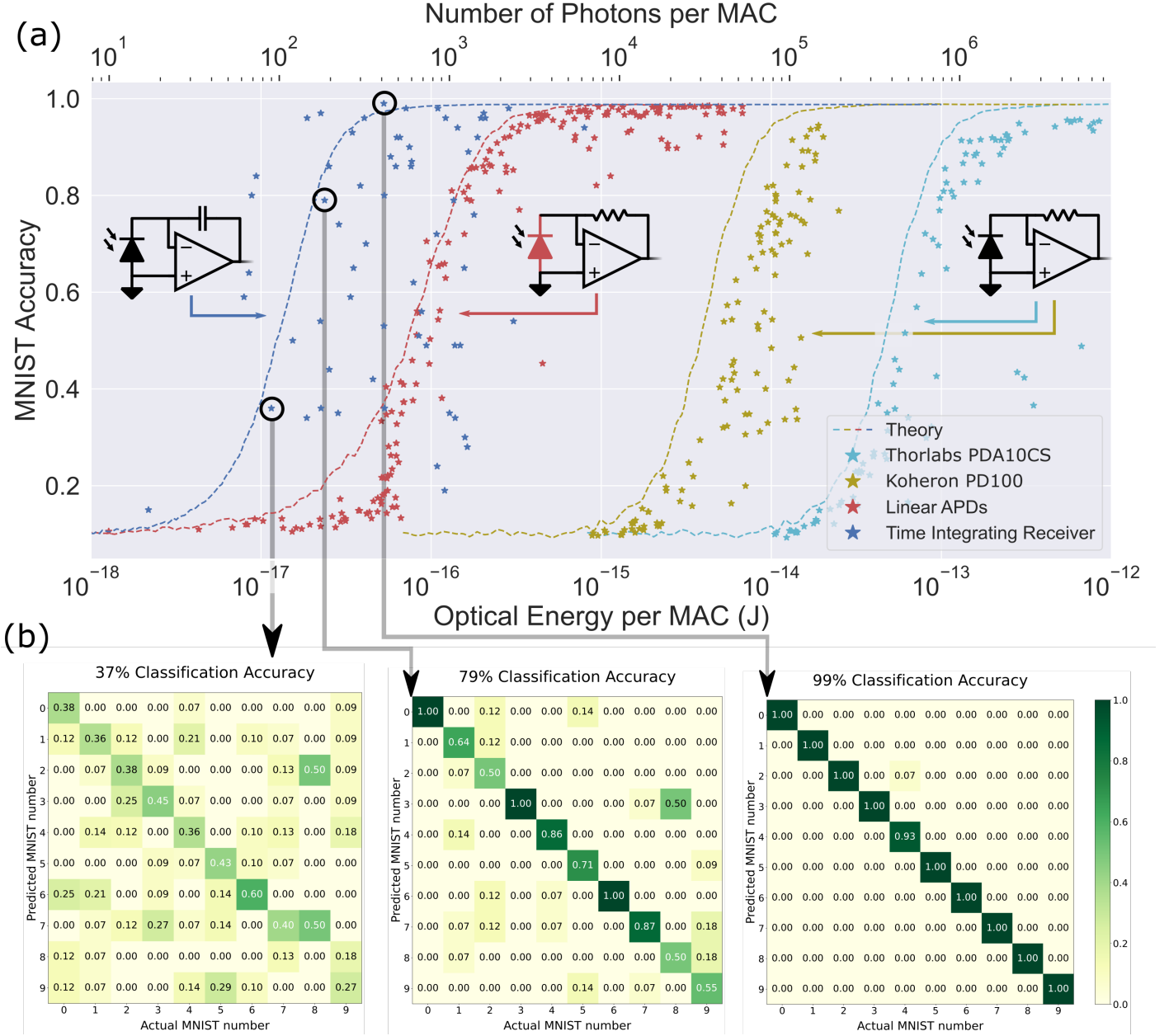

Figure 4. Thermal noise limited optical sensitivity of Netcast system. (a) Experimentally measured sensitivity of optical receivers. Standard amplified photoreceivers are shown on the right side of the plot, with performance limited by electrical amplifier thermal noise, giving a typically optical energy of 10-100 fJ/MAC. The center of the plot shows linear avalanche photodiodes, which use intrinsic gain to lower the energy per MAC, but at the cost of increased energy consumption and lower bandwidth time integrating receivers, which lower the effective thermal noise floor by performing many MAC operations for each readout. Time integrating receivers utilizing off-the-shelf technology can achieve high accuracy with <100 aJ/MAC of optical sensitivity on the benchmark neural network task. (b) Confusion matrices for labeled points in subfigure (a) showing how each digit in the MNIST dataset is classified by the optical hardware (on-diagonal elements correspond to correct classification; columns add to 1 but rows do not have to).

We investigate this fundamental bound of the Netcast system by using superconducting nanowire single-photon detectors (SNSPDs) as shown in Fig. 5(b). These photodetectors are ideal, demonstrating pure shot-noise limited performance. We show that the fundamental shot noise bound on the same benchmark digit classification problem from Fig. 4 allows the receiver to operate with high accuracy with < 1 photon per MAC (0.1 aJ/MAC). This result may at first seem surprising since having less than a single photon per MAC is counterintuitive. We can understand this measurement better by noting that at readout we have performed a vector-vector product with $M = 100$ MACs. Each MAC can have less than a single photon in it, but the measured signal will have many photons in it. A graphical explanation of this is in Supplementary Materials XVIII. This single photon per MAC regime enables many new applications. For security, deployed machine learning models are now secure to eavesdroppers trying to learn model weights.

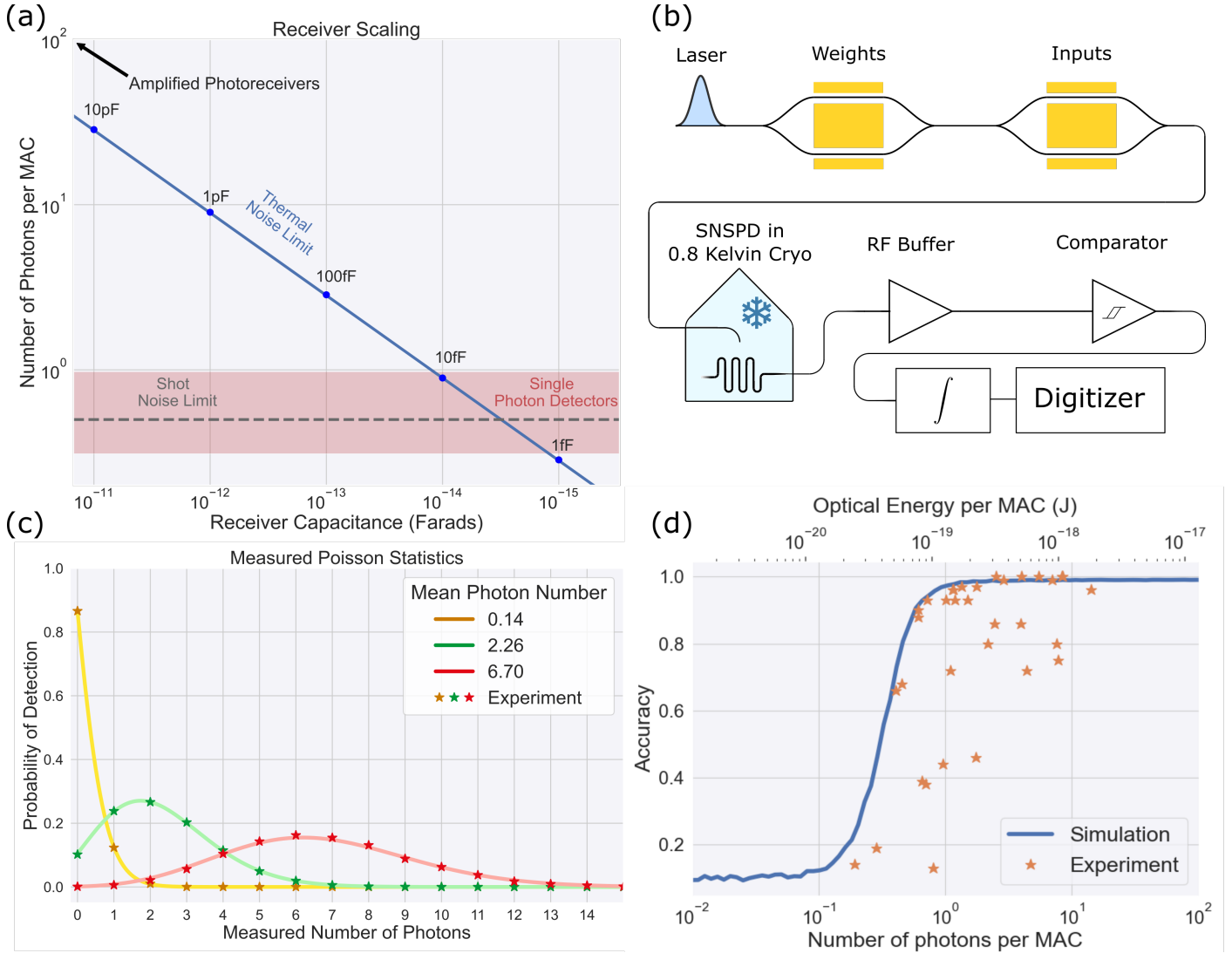

Figure 5. Forward looking performance of Netcast. (a) Fundamental noise bounds of time-integrating receivers from thermal noise of an integrator and shot noise to achieve 50% accuracy on MNIST task. Decreasing the capacitance of the time integrator lowers thermal readout noise, enabling access to the single photon per MAC regime. (b) Experimental setup consisting of input and weight modulators and superconducting nanowire single photon detectors (SNSPDs), allowing us to probe this fundamental single-photon bound. (c) We experimentally validate the single photon detectors by measuring shot-noise on the detector over many integration windows. (d) Using a 3-layer MNIST model we experimentally measure computation with <1 photon per MAC with high accuracy.

With less than one photon per weight, weight data which is distributed to the client can be secured by the laws of quantum mechanics. An eavesdropper measuring deployed weight data over the fiber would not be able to learn the values of individual weights, but is restricted to only their mean statistics. This enables accurate computation where both an eavesdropper and the client are blind to the weight data being received. Companies that spend significant capital creating models for inference can use this technique to perform computation away from their cloud hardware without anyone stealing the model. Another application which benefits from less than one photon per MAC is deployed spacecraft that operate in a strongly photon starved environment. Weight data from a directional base station could be transmitted to the spacecraft and classified on the craft, before the results are transmitted to earth.

## DISCUSSION

The system level demonstration shown here is one example of an implementation of Netcast. The cloud-based smart transceiver proposed in this paper can reside inside of existing networking hardware such as network switches, servers, or edge nodes. Our ideas can be extended to the case where the user data is streamed through programmable network switches with smart transceivers, enabling in-network optical inference [18]. Modern network switches are an ideal platform for developing Netcast commercially, as they are programmable, enabling multiple streams of weights to be deployed at line rate (100Gbps) and can support 64GB of memory, exceeding the storage requirements of modern neural networks [33, 34]. Prior work has demonstrated the feasibility of using programmable switches to perform layer-by-layer inference with smart transceivers [18]. The large data storage of these network switches enables multiple models to be stored and queried. The client device could use it's broadband modulator to allow for reflection-mode communication back to the server, where the client modulates received light and sends it back along the fiber link for communication. This querying communication can be slow and lossy as only a few bits are required to request a new model be sent.

Emerging photonic technologies, such as low power static phase shifters [35–38] and high-speed phase shifters [39–42], can reduce receiver electrical energy consumption to ≈10 aJ/MAC. This energy can be further decreased by making use of the tight integration of transistors and photonics in silicon using technologies such as receiverless detectors [11], photonic DACs [43], and photonic ADCs [44]. Detectors such as avalanche detectors could be incorporated with a time-integrator to provide a benefit to the optical sensitivity of the receiver, but at the cost of added electrical power consumption (Supplementary Materials XIX). Further improvements in optical sensitivity are possible by utilizing coherent detection, which boosts the received signal using a strong local oscillator [27]. Detailed in Supplementary Materials XII are two examples of a Netcast system using coherent detection to substantially improve optical energy per MAC.

A number of companies have designed custom edge computing ASICs with reduced SWaP [7, 45], but these ASICs are hampered by the same energy and bandwidth constraints of larger CMOS processors. Analog accelerators, such as memristive crossbar arrays and meshes of photonic interferometers, hold promise for lowering the power consumption of neural networks compared to electronic counterparts, but existing commercial demonstrations still consume watts of power [8, 9].

One obstacle to scaling bandwidth in traditional optical communication systems is dispersion in optical fiber. For a single smart transceiver and client, techniques such as wavelength dependent delays can compensate for dispersion at the smart transceiver. However, in systems where weights are deployed to multiple clients from one smart transceiver with different lengths of fiber, this technique cannot be used. We discuss the effects of dispersion in Supplementary Material XX and show that it is possible to make use of the optical O-band to enable THz of bandwidth at clock rates of 10 GHz per wavelength over more than 10 km of optical fiber.

## CONCLUSION

We have described a novel edge computing architecture that makes use of the strengths of photonics and electronics to achieve orders of magnitude of improvement over existing digital electronics. We have demonstrated scalable photonic edge computing utilizing WDM, time-integrating receivers, scalability to <1 photon per MAC and computing over deployed fiber using 3 THz of bandwidth. On these tasks, we show 98.8% accurate image classification. The hardware shown in this paper is readily mass-producible from existing CMOS foundries, allowing for near-term impact on our daily lives. Our approach removes a fundamental bottleneck in computing, enabling life-changing applications on the internet's peripheral nervous system ranging from high-speed computing on deployed sensors and drones, live video processing on cellular devices and networks and possibly image classification on spacecraft searching for life at the edge of the solar system.

## METHODS

### Silicon Smart Transceiver

The smart transceiver is composed of 48 silicon photonic MZMs each capable of modulation at 50 Gbps. Each MZM consists of two thermo-optic phase shifters for controlling the bias point of the modulator and two high-speed free-carrier plasma dispersion phase shifters to control the optical intensity transmitted [46]. The chip occupies 422 mm$^2$ and is connected to a printed circuit board for testing through 336 wirebonds. To couple light in and out of the chip we align a 64 channel polarization maintaining fiber array to 62 grating couplers. A bank of lasers with distinct wavelengths (Optilab TWL-4-B-MIC) are coupled into the smart transceiver and individually modulated. We use electrical arbitrary waveform generators (Keysight M3202A) to map floating point weight and input values onto the transmitted optical intensity. The modulated outputs are then multiplexed onto a signal deployed over optical fiber.

The 48 channel silicon smart transceiver was fabricated in the OpSIS IME foundry process multi-project wafer run. The Institute for Microelectronics (IME) foundry (now called AMF) uses 248 nm lithography and is capable of producing 130 nm CMOS electronics. The smart transceiver is designed for parallel modulation of light in the optical C-band (1550 nm). This Silicon-on-Insulator (SOI) process creates 220nm thick silicon components with 2 $\mu$m of buried oxide below them and includes photodetectors, multiple doping layers, metal layers, and partial etch layers. More details on the OpSIS IME process can be found in [47]. 500 nm fully-etched waveguides were used for short-distance light routing with 1.2 um wide waveguides used for long-distance routing to reduce propagation losses. Further details on the smart transceiver can be found in [19].

### Optical Energy Efficiency Measurement

To generate the data shown in Fig. 4 and Fig. 5 several photoreceivers are used and benchmarked. Details of the photodetectors used and their calibration methods can be found in Supplementary Materials IV. The superconducting nanowire single photon detectors were calibrated by measuring the voltage at the output of the integrator after a fixed integration window. As shown in Supplemental Materials XVII these voltages form distinct "bins", enabling us to map a measured voltage to a number of photons. This mapping is used to create the Poisson noise statistics measured in Fig. 5, verifying its accuracy.

## ACKNOWLEDGEMENTS

This research is funded by a collaboration with NTT-Research and NSF Eager (CNS-1946976). This material is based on research sponsored by the Air Force Office of Scientific Research (AFOSR) under award number FA9550-20-1-0113, the Air Force Research Laboratory (AFRL) under agreement number FA8750-20-2-1007, the Army Research Office (ARO) under agreement number W911NF-17-1-0527, NSF RAISE-TAQS grant number 1936314 and NSF C-Accel grant number 2040695. Distribution Statement A. Approved for public release. Distribution is unlimited. This material is based upon work supported by the Under Secretary of Defense for Research and Engineering under Air Force Contract No. FA8702-15-D-0001. Any opinions, findings, conclusions or recommendations expressed in this material are those of the authors and do not necessarily reflect the views of the Under Secretary of Defense for Research and Engineering.

A.S and S.B are supported by a National Science Foundation Graduate Research Fellowship 1745302.

The author would like to acknowledge David Lewis and Anthony Pennes for assistance in machining laboratory equipment and Euan Allen for discussions related to using squeezed light to further reduce photon counts. We are grateful to Eric Bersin and Ben Dixon for assistance in co-ordinating usage of the deployed fiber and Anthony Rizzo for his help in converting and plotting eye diagram data as well as proof reading the manuscript. We would like to thank Christopher Panuski and Stefan Krastanov for informative discussions on single photon operation of Netcast. We thank Adrian Pyke of Micro Precision Technologies for wire bonding the electrical connections from the printed circuit board to the 48-channel transmitter. We appreciate help from Franco Wong for the use of his SNSPDs and acknowledge Mihika Prabhu and Carlos Errando Herranz for facilitating the usage of the SNSPDs, Charles Freeman for his help in taking drone photography, Nvidia for supplying a Tesla K40 GPU which was used for simulations shown in the main text and supplemental, and Planet.org for allowing us to take custom satellite imagery.

## AUTHOR CONTRIBUTIONS

A.S created the experimental setup and conducted the experiment. S.B assisted in fiber-to-chip coupling and discussions on the project. Z.C assisted with high-speed measurements of the setup and discussions. M.S, A.N, T.B.J, and M.H designed and taped out the smart transceiver. D.B packaged the 48-channel silicon transceiver. J.C packaged the time-integrating receivers and assisted in calibration. D.E established the fiber link between MIT and MIT Lincoln Laboratory and S.H and P.B.D helped with its use. L.B helped in discussion of fundamental noise sources. M.G and Z.Z assisted with discussions on modern telecommunication networks. R.H, D.E and M.G conceived of the project idea. L.B, M.S, A.N, T.B.J, M.H, Z.Z, J.C, S.H, P.B.D and M.G provided feedback on the manuscript. A.S, D.E, and R.H wrote the manuscript.

## COMPETING INTERESTS

M.S is the leader of silicon photonics at Nokia Corporation. M.H is president of Luminous computing. A.N is a system architect at Luminous computing. T.B.J is Vice-President of engineering at Luminous computing. D.B is Chief Scientist at Lightmatter. R.H and D.E have filed a patent related to Netcast: PCT/US21/43593. M.G, Z.Z, L.B, A.S, R.H, D.E have filed a provisional patent related to Netcast: 63/191,120 Other authors declare no competing interests.

## DATA AVAILABILITY

The data supporting the claims in this paper is available upon reasonable request.

## CORRESPONDENCE

Requests for information should be directed to Alexander Sludds (asludds@mit.edu).

---

[1] T. B. Brown, B. Mann, N. Ryder, M. Subbiah, J. Kaplan, P. Dhariwal, A. Neelakantan, P. Shyam, G. Sastry, A. Askell, *et al.*, arXiv preprint arXiv:2005.14165 (2020).

[2] O. Vinyals, I. Babuschkin, W. M. Czarnecki, M. Mathieu, A. Dudzik, J. Chung, D. H. Choi, R. Powell, T. Ewalds, P. Georgiev, *et al.*, Nature **575**, 350 (2019).

[3] A. Krizhevsky, I. Sutskever, and G. E. Hinton, Advances in neural information processing systems **25**, 1097 (2012).

[4] J. Deng, W. Dong, R. Socher, L.-J. Li, K. Li, and L. Fei-Fei, in *2009 IEEE conference on computer vision and pattern recognition* (Ieee, 2009) pp. 248–255.

[5] V. Sze, Y.-H. Chen, T.-J. Yang, and J. S. Emer, Proceedings of the IEEE **105**, 2295 (2017).

[6] M. Davies, N. Srinivasa, T.-H. Lin, G. Chinya, Y. Cao, S. H. Choday, G. Dimou, P. Joshi, N. Imam, S. Jain, *et al.*, Ieee Micro **38**, 82 (2018).

[7] Mythic ai analog matrix processor (2022), accessed: 2022-02-01.

[8] C. Demirkiran, F. Eris, G. Wang, J. Elmhurst, N. Moore, N. C. Harris, A. Basumallik, V. J. Reddi, A. Joshi, and D. Bunandar, arXiv preprint arXiv:2109.01126 (2021).

[9] D. Fick and M. Henry, Hot Chips 2018 (2018).

[10] M. Horowitz, in *2014 IEEE International Solid-State Circuits Conference Digest of Technical Papers (ISSCC)* (IEEE, 2014) pp. 10–14.

[11] D. A. Miller, Journal of Lightwave Technology **35**, 346 (2017).

[12] L. Bernstein, A. Sludds, R. Hamerly, V. Sze, J. Emer, and D. Englund, Scientific reports **11**, 1 (2021).

[13] Y.-H. Chen, T. Krishna, J. S. Emer, and V. Sze, IEEE journal of solid-state circuits **52**, 127 (2016).

[14] M. Satyanarayanan, Computer **50**, 30 (2017).

[15] R. Mac, New York Times (2021).

[16] E. Bowman, National Public Radio (2021).

[17] R. Hamerly, A. Sludds, S. Bandyopadhyay, L. Bernstein, Z. Chen, M. Ghobadi, and D. Englund, in *Emerging Topics in Artificial Intelligence (ETAI) 2021*, Vol. 11804 (International Society for Optics and Photonics, 2021) p. 118041R.

[18] Z. Zhong, W. Wang, M. Ghobadi, A. Sludds, R. Hamerly, L. Bernstein, and D. Englund, in *Proceedings of the ACM SIGCOMM 2021 Workshop on Optical Systems* (2021) pp. 18–22.

[19] M. Streshinsky, A. Novack, R. Ding, Y. Liu, A. E.-J. Lim, P. G.-Q. Lo, T. Baehr-Jones, and M. Hochberg, Journal of Lightwave Technology **32**, 4370 (2014).

[20] T. Gokmen, M. J. Rasch, and W. Haensch, in *2019 IEEE International Electron Devices Meeting (IEDM)* (IEEE, 2019) pp. 22–3.

[21] S. Garg, J. Lou, A. Jain, and M. Nahmias, arXiv preprint arXiv:2102.06365 (2021).

[22] B. M. Pietro Caragiulo, Clayton Daigle, Dac performance survey 1996-2020 (2022).

[23] B. Murmann, Adc performance survey 1997-2021 (2022).

[24] E. Yang and T. Lehmann, ISCAS 2019 (2019).

[25] A. N. Tait, arXiv preprint arXiv:2108.04819 (2021).

[26] J. B. Johnson, Physical review **32**, 97 (1928).

[27] R. Hamerly, L. Bernstein, A. Sludds, M. Soljačić, and D. Englund, Physical Review X **9**, 021032 (2019).

[28] T. Wang, S.-Y. Ma, L. G. Wright, T. Onodera, B. C. Richard, and P. L. McMahon, Nature Communications **13**, 1 (2022).

[29] J. Pierce, Proceedings of the IRE **44**, 601 (1956).

[30] M. Rakowski, C. Meagher, K. Nummy, A. Aboketaf, J. Ayala, Y. Bian, B. Harris, K. Mclean, K. McStay, A. Sahin, *et al.*, in *Optical Fiber Communication Conference* (Optical Society of America, 2020) pp. T3H–3.

[31] C. Sun, M. Wade, M. Georgas, S. Lin, L. Alloatti, B. Moss, R. Kumar, A. H. Atabaki, F. Pavanello, J. M. Shainline, *et al.*, IEEE Journal of Solid-State Circuits **51**, 893 (2016).

[32] N. Mehta, Z. Su, E. Timurdogan, J. Notaros, R. Wilcox, C. Poulton, C. Baiocco, N. Fahrenkopf, S. Kruger, T. Ngai, *et al.*, in *2020 IEEE Symposium on VLSI Technology* (IEEE, 2020) pp. 1–2.

[33] Intel tofino 3 intelligent fabric processor brief.

[34] Juniper networks: Mx series universal routing platform.

[35] K. Giewont, K. Nummy, F. A. Anderson, J. Ayala, T. Barwicz, Y. Bian, K. K. Dezfulian, D. M. Gill, T. Houghton, S. Hu, *et al.*, IEEE Journal of Selected Topics in Quantum Electronics **25**, 1 (2019).

[36] G. Liang, H. Huang, A. Mohanty, M. C. Shin, X. Ji, M. J. Carter, S. Shrestha, M. Lipson, and N. Yu, Nature Photonics **15**, 908 (2021).

[37] R. Baghdadi, M. Gould, S. Gupta, M. Tymchenko, D. Bunandar, C. Ramey, and N. C. Harris, Optics Express **29**, 19113 (2021).

[38] M. Dong, G. Clark, A. J. Leenheer, M. Zimmermann, D. Dominguez, A. J. Menssen, D. Heim, G. Gilbert, D. Englund, and M. Eichenfield, Nature Photonics **16**, 59 (2022).

[39] E. Timurdogan, C. M. Sorace-Agaskar, J. Sun, E. S. Hosseini, A. Biberman, and M. R. Watts, Nature communications **5**, 1 (2014).

[40] C. Wang, M. Zhang, X. Chen, M. Bertrand, A. Shams-Ansari, S. Chandrasekhar, P. Winzer, and M. Lončar, Nature **562**, 101 (2018).

[41] M. Xu, Y. Zhu, F. Pittalà, J. Tang, M. He, W. C. Ng, J. Wang, Z. Ruan, X. Tang, M. Kuschnerov, *et al.*, Optica **9**, 61 (2022).

[42] W. Heni, Y. Fedoryshyn, B. Baeuerle, A. Josten, C. B. Hoessbacher, A. Messner, C. Haffner, T. Watanabe, Y. Salamin, U. Koch, *et al.*, Nature communications **10**, 1 (2019).

[43] S. Moazeni, S. Lin, M. Wade, L. Alloatti, R. J. Ram, M. Popović, and V. Stojanović, IEEE Journal of Solid-State Circuits **52**, 3503 (2017).

[44] A. Zazzi, J. Müller, M. Weizel, J. Koch, D. Fang, A. Moscoso-Mártir, A. T. Mashayekh, A. D. Das, D. Drayß, F. Merget, *et al.*, IEEE Open Journal of the Solid-State Circuits Society **1**, 209 (2021).

[45] A. Yazdanbakhsh, K. Seshadri, B. Akin, J. Laudon, and R. Narayanaswami, arXiv preprint arXiv:2102.10423 (2021).

[46] R. Soref and B. Bennett, IEEE journal of quantum electronics **23**, 123 (1987).

[47] T. Baehr-Jones, R. Ding, A. Ayazi, T. Pinguet, M. Streshinsky, N. Harris, J. Li, L. He, M. Gould, Y. Zhang, *et al.*, arXiv preprint arXiv:1203.0767 (2012).

# Supplementary Material: Delocalized Photonic Deep Learning on the Internet's Edge

Alexander Sludds[1], Saumil Bandyopadhyay[1], Zaijun Chen[1], Zhizhen Zhong[2], Jared Cochrane[1,3], Liane Bernstein[1], Darius Bunandar[1], P. Ben Dixon[3], Scott A. Hamilton[3], Matthew Streshinsky[4], Ari Novack[4], Tom Baehr-Jones[4], Michael Hochberg[4], Manya Ghobadi[2], Ryan Hamerly[1,5], and Dirk Englund[1]

[1]*Research Laboratory of Electronics, MIT, Cambridge, MA, 02139, USA*

[2]*Computer Science and Artificial Intelligence Laboratory, MIT, Cambridge, MA, 02139, USA*

[3]*Lincoln Laboratory, MIT, Lexington, MA, 02421, USA*

[4]*Nokia Corporation, 171 Madison Avenue Ste 1100, New York, NY, 10016, USA and*

[5]*NTT Research Inc., PHI Laboratories, 940 Stewart Drive, Sunnyvale, CA 94085, USA*

## CONTENTS

# I. EXPERIMENTAL SETUP

Supplementary Figure 1 shows the system scale packaging and control for the smart transceiver, including optical access through a 64 channel, 8-degree polished, 127 um pitch, polarization maintaining fiber array (custom from PM optics) and a custom 64 channel fiber patch panel. The setup is optically driven by a bank of 16 tunable lasers (Optilab TWL-4-B-MIC) and a 16 channel 1 gigasample per second arbitrary waveform generator (Keysight M3202 inside a Keysight M9010A chasis). The smart transceiver is thermally stabilized by a temperature controller (Arroyo Instruments 5240).

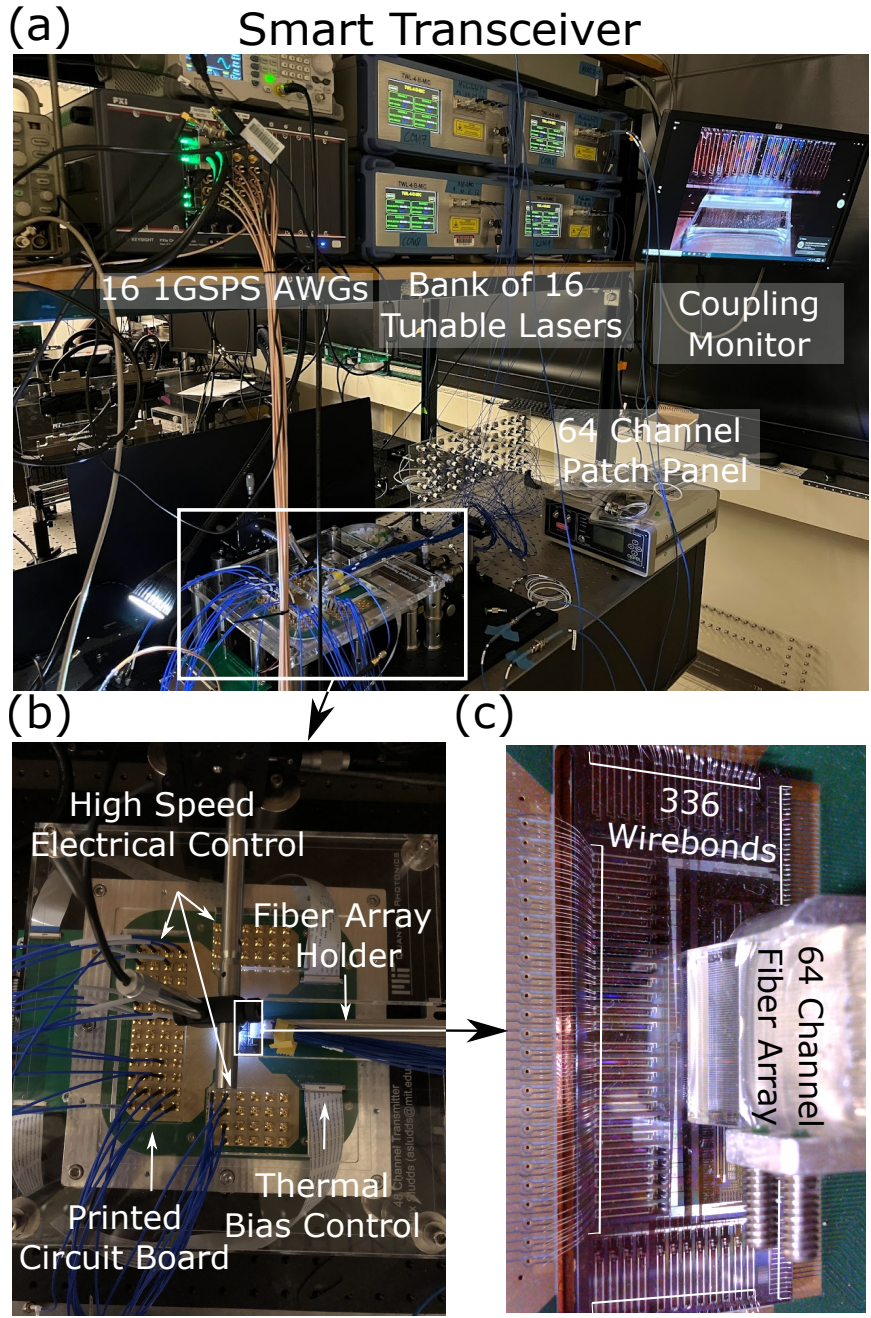

Supplementary Figure 1: Smart Transceiver Experimental Setup. (a) Zoomed out view of smart transceiver setup showing 16 1 Gigasample per second (GSPS) arbitrary waveform generator channels (AWGs), a bank of 16 tunable C-band lasers, and a 64 channel fiber patch panel. (b) Zoom in on smart transceiver packaging composed of a printed circuit board, fiber array holder, array of 96 high-speed electrical control connectors and two ribbon cables for thermal bias control of 96 thermal phase shifters. (c) Zoom in of the 442 mm² silicon smart transceiver with 336 wirebonds and 64 channel fiber array highlighted.

The client, shown in Supplementary Figure 2, is composed of an input modulator (Lithium Niobate, JSDU AM-150, 20 GHz bandwidth), wavelength division multiplexer (WDM, custom from Fiberdyne), bias controller for input modulator (NIDAQ BNC6343), and integrators. Integrators are composed of an InGaAs photodiode (Thorlabs FGA01FC) and integrating IC (Texas Instruments IVC102). Readout is performed with 2x Spectrum Instrument M2p.5943-x4. The left BNC port of the integrator board is for triggering/resetting the integrator and the right 4 ports are for electrical readout. Supplementary Figure 2(d) shows the layout of this integrating IC, which was revealed by de-encapsulating the packaged IC in a 98% concentrated solution of 200 C nitric acid for 5 minutes, followed by 10 minutes of sonication in acetone.

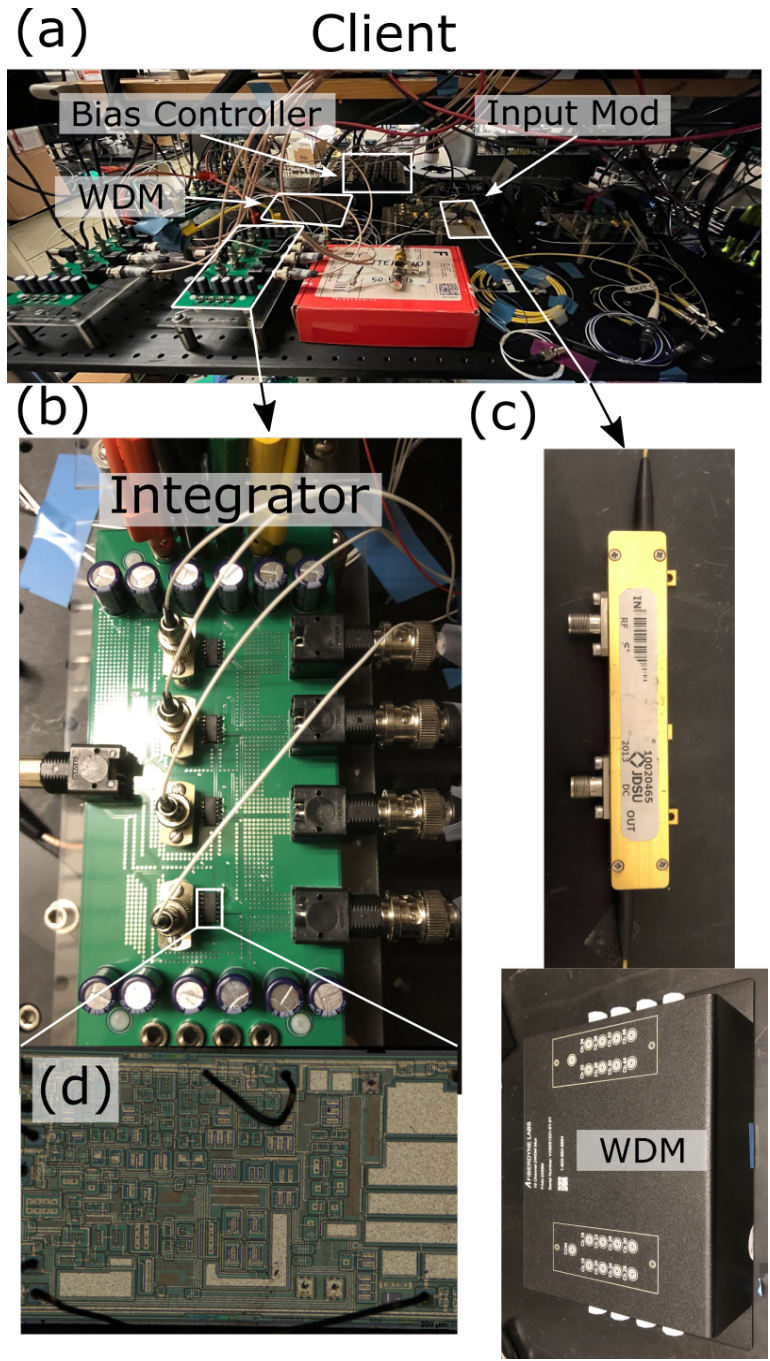

Supplementary Figure 2: Client Experimental Setup. (a) Zoomed out view of client setup including input modulator, bias controller for the input modulator, wavelength division multiplexer (WDM), and a bank of 16 time integrating receivers with 4 integrators per PCB. (b) The time integrator board is composed of 4 pairs of photodiodes and integrators. (c) Input modulator and WDM output of the setup. (d) Integrator IC after nitric acid delidding.

## II. SILICON PHOTONIC COMPONENTS

The demonstrations performed in the main text are done using a silicon photonic smart transceiver, the components of which are described in Supplementary Figure 3. Each modulator is composed of high-speed traveling wave modulators and efficient thermo-optic modulators in silicon. The design of the former is discussed in the reference text, though the wavelength we use is in the C band (1550 nm) rather than the O band (1310 nm) [1]. These modulators have a traveling wave impedance of $\approx 33$ Ohms and have a $V_\pi L$ of 2.5 V-cm. Recent advances in silicon photonic modulator design have enabled Mach-Zehnder modulators with higher-bandwidth (50 GHz), loss below 3 dB and $V_\pi L$ values of 0.45 V-cm with shorter lengths [2]. We note here that the schematic in the figure for the modulator is in a ground-signal-ground (GSG) configuration. In practice silicon carrier modulators are GSGSG, with separate electrodes for each arm of the Mach-Zehnder. This is done to avoid forward biasing the PN junction.

The thermal phase shifters are made by doping silicon near the optical waveguide. By running current through this doped silicon a change in temperature is created through Joule heating. This "heater" section of silicon uses this change in temperature to shift the phase of light using silicon's thermo-optic effect [3]. The measured resistance of these thermal phase shifters in the lab was $\approx 185$ Ohms, and their measured $P_\pi$ was 90 mW. Thermal phase shifters in the same platform have shown $P_\pi$ of 25 mW without process modifications [4]. This value can be further lowered to 1 mW by under-etching the phase shifter or to zero hold power by using mechanical phase shifters [5, 6].

The grating couplers in the weight server have been shown to have an insertion loss of 4.4 dB per facet, leading to $\approx 9$ dB of through chip loss. We also see an additional 6 dB of loss from using y-splitters on chip to combine sets of 4 modulator outputs together onto 1 output grating each.

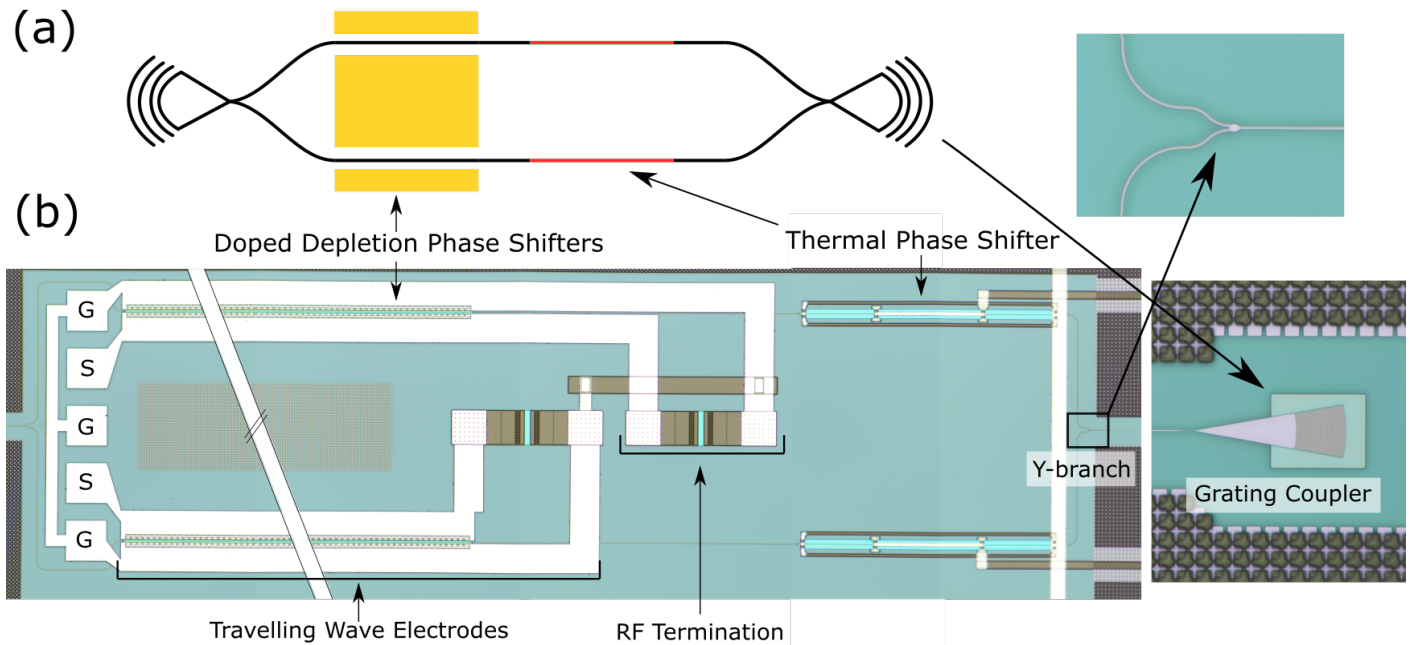

Supplementary Figure 3: (a) Schematic of components used on the silicon photonic smart transceiver include (a) gratings, electro-optic modulators, thermo-optic modulators, and y-branches. (b) The electro-optic modulator is experimentally implemented as a traveling-wave silicon photonic carrier depletion Mach-Zehnder modulator terminated by doped RF terminations. Thermal phase shifters enable bias control of the modulator. Passive silicon Y-branches allow for splitting and combining of light. Grating couplers allow for fiber-coupling of light to the smart transceiver.

## III. SINGLE PHOTON DETECTION EXPERIMENTAL SETUP

In the main text we probe the limits of detection in Netcast by utilizing single photon detectors (superconducting nanowire single photon detectors, SNSPDs) as ideal shot noise limited detectors. The experimental setup is described in Supplementary Figure 4. We utilize two fiber coupled lithium niobate modulators (JDSU AM-150) to control weight and input values. These feed into an SNSPD in a cryo (Tungsten Silicide (WSi) SNSPDs fabricated by NIST, similar to detectors found in [7] and the same detectors as found in [8, 9]). We use ≈30 meters of RF cable to connect the cryo in one room to the integrator in a separate room and buffer the long RF cable with an RF buffer amplifier (Femto 200MHz DHPVA configured in 30 dB mode). A comparator, composed of a Teensy 4.0 microcontroller, is then used to convert each SNSPD pulse into a 1 us wide rectangular pulse. A high-speed line buffer (Thorlabs 50LD) buffers the signal after the comparator. A stack of RF attenuators totalling 31 dB attenuation decreases the voltage for the voltage integrator. A custom voltage integrator PCB, composed of an IVC102 IC utilizing the 10 pF internal integration capacitance with a 300 kOhm series resistor is used to integrate the voltage from the Teensy. A DAQ controller (NIDAQ 6343) is used for data encoding and readout.

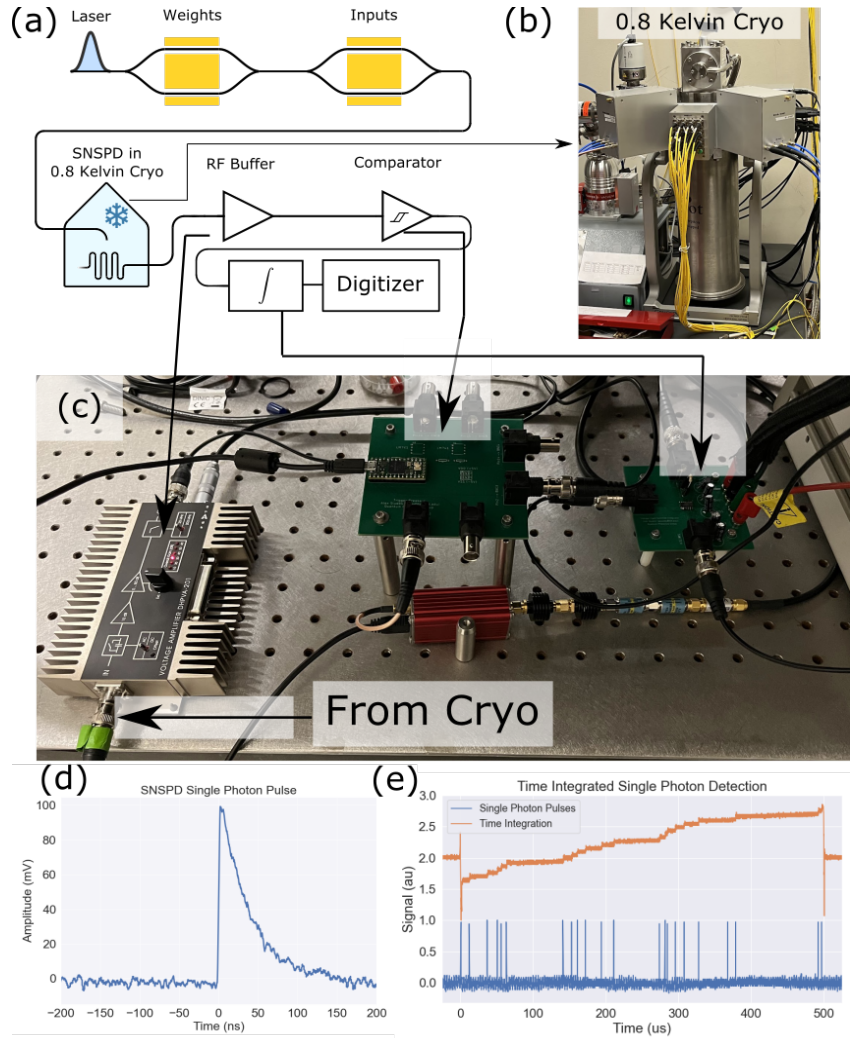

Supplementary Figure 4: (a) Schematic showing the experimental setup for measuring less than a single photon per MAC. (b) SNSPD in a 0.8 Kelvin cryostat. (c) The generated electrical voltage pulses go to a custom electrical subsystem consisting of an RF buffer, electrical comparator, voltage integrator, and digitizer.(d) An example electrical pulse from the SNSPD before the RF buffer. (e) An example of a time-integrated waveform alongside SNSPD pulses. The integration waveform is offset vertically for visibility.

## IV. CALCULATION OF RECEIVER NOISE

Five different photodetectors are shown in the main text:

  A Amplified detector 1: Thorlabs PDA10CS

  B Amplified detector 2: Koheron PD100

  C Amplified linear avalanche photodetector: Thorlabs APD430C

  D Custom time integrating receiver (described below)

  E Superconducting nanowire single photon detector

Below is a derivation of the noise calculation used for Figure 4 of the main text. Throughout this section we will be defining signal to noise ratio as the ratio of received optical power to the dominant noise source. This can be converted to units of readout voltage and quoted as $\mathrm{SNR} = \frac{V_{\mathrm{out}}}{V_{\mathrm{rms}}}$ for consistency.

### A. Thorlabs PDA10CS

The Thorlabs PDA10CS IR photodetector has several gain settings, allowing the user to tradeoff analog bandwidth for electrical gain. We use the 30 dB gain setting, which gives the best performance for the photoreceiver. Driving a 50 ohm load, the RMS noise of the receiver is $V_{rms} = 300~\mu$V with a bandwidth of $B = 775$kHz and a gain of $G = 2.4 \times 10^4$ V/A. The IR responsivity of the detector is $\eta \approx 1$ A/W. The signal-to-noise ratio for a given energy per MAC of this detector can be calculated as

$$\mathrm{SNR} = \frac{\mathrm{P_{opt}}}{\mathrm{NEP}} = \frac{\mathrm{E_{mac} \cdot B}}{\frac{V_{\mathrm{rms}}}{\eta G}} \tag{1}$$

where $P_{opt}$ is the optical power incident on the receiver and NEP is the integrated noise equivalent power in units of watts.

We then feed this signal to noise ratio into a software model which adds noise approximating this signal to noise ratio to the model. The result of this is plotted as dashed lines in Figure 4.

Supplementary Figure 5 shows the noise spectrum of the PDA10CS detector as well as the quoted noise equivelent power. From this spectrum an integrated voltage noise of $270\mu$V is estimated, in close agreement with the quoted $300\mu$V from the manufacturer. For this supplementary section a Rigol DSA815 RF spectrum analyzer is used to measure spectrum data with its RF preamplifier option enabled. A 1 kHz sampling bandwidth is used and the measured data is then normalized and converted to units of NEP using the specified gain from the specification sheet.

### B. Koheron PD100-DC

The Koheron PD100-DC is an InGaAs amplified photodetector with a fixed transimpedance gain of $G = 3900\frac{V}{A}$, a responsivity of $0.9\frac{A}{W}$ and a bandwidth of $B = 110$MHz. The integrated noise of this detector is calculated using the input current density noise from the datasheet $7\mathrm{pA}/\sqrt{\mathrm{Hz}}$ and converted to an integrated voltage noise of $286~\mu$V by multiplying by the square root of bandwidth and the transimpedance gain. The signal to noise ratio is then calculated in the same manner as the Thorlabs PDA10CS, and the same modeling technique is used. An additional factor of 2 is taken into account from the measured voltage signal in the experiment to compensate for a 50 Ohm termination on the digitizer.

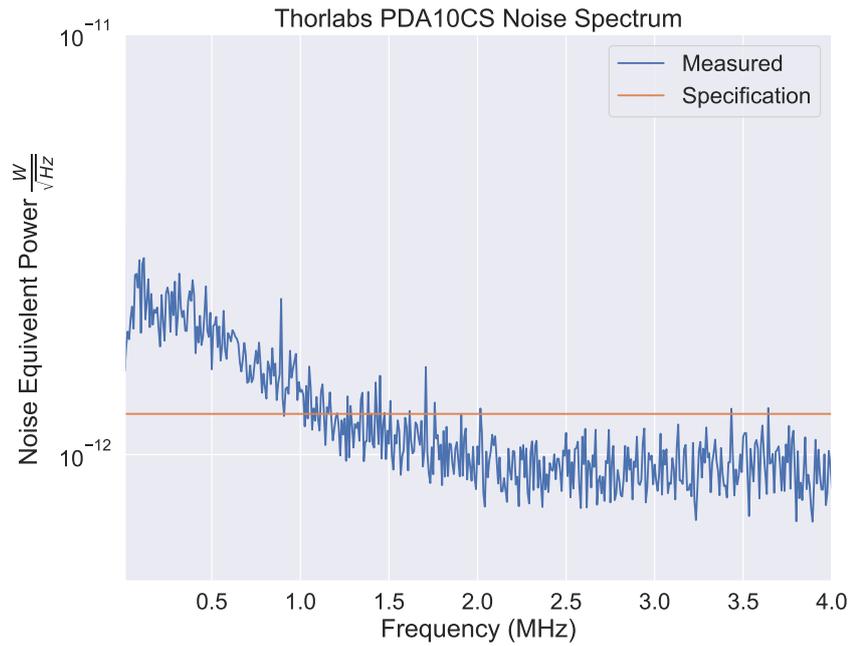

Supplementary Figure 5: Noise spectrum of Thorlabs PDA10CS Amplified Photodetector

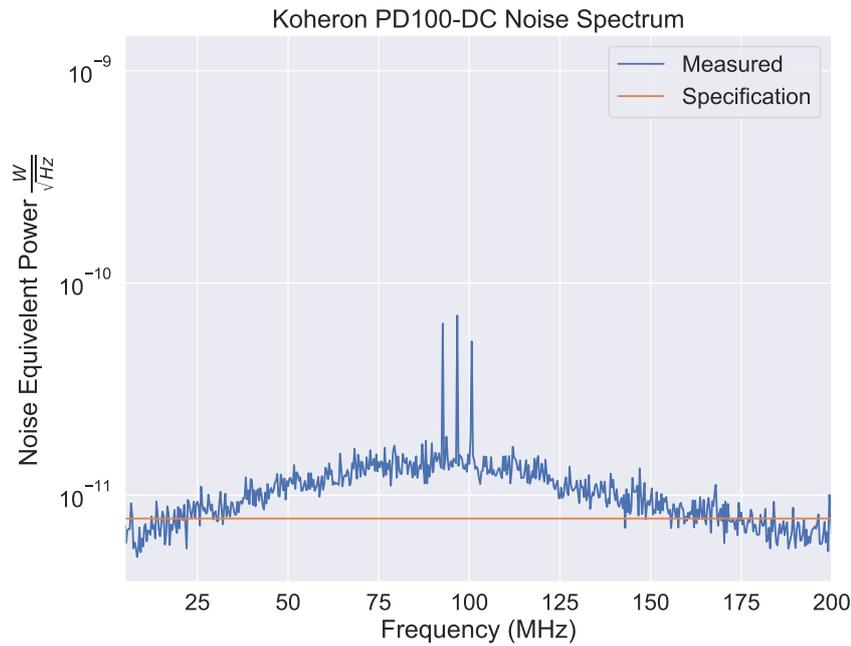

Supplementary Figure 6: Noise spectra of Koheron PD100-DC Amplified Photodetector

Supplementary Figure 6 shows the noise spectrum of the Koheron PD100-DC amplified photodetector.

### C. Thorlabs APD430C

The Thorlabs APD430C is a linear mode avalanche photodetector with an internal gain of $M = 20$. In this mode the conversion gain is $G = 1.8 \cdot 10^5 \frac{V}{W}$, bandwidth is $B = 400$ MHz, responsivity of $\eta = 0.9 \frac{A}{W}$ and integrated optical noise is 17 nW. This noise is converted to a voltage noise of $V_{rms} = 3$ mV by using the gain above. The calculation of SNR continues as in the above sections.

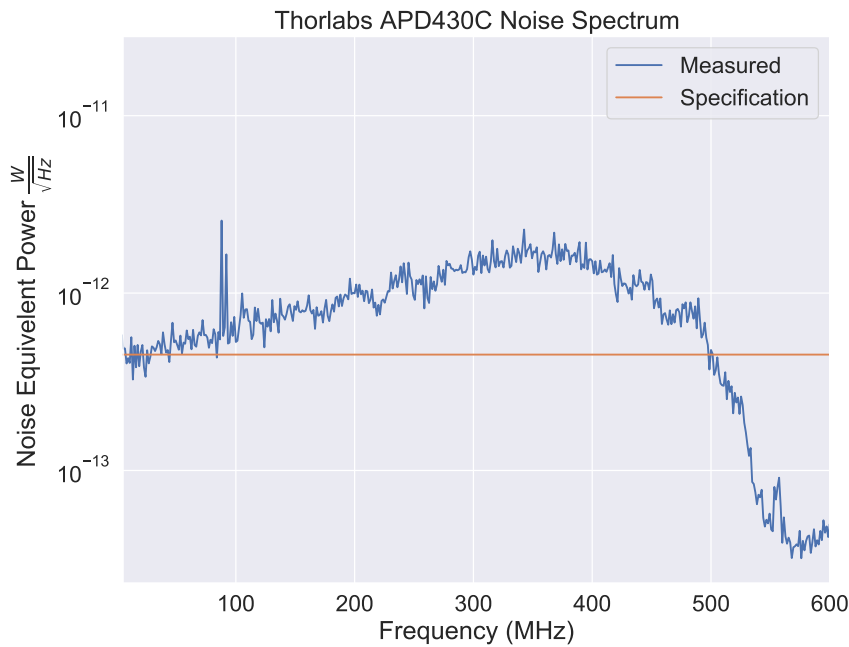

Supplementary Figure 7: Noise spectra of Thorlabs APD430C Amplified Linear Mode Avalanche Photodetector

In the experimental data we only had a 50% coupling efficiency to the photodetector because of the available fiber faceplates for the detector. This factor of 2 is taken into account in the simulations and text.

Supplementary Figure 7 shows the noise spectrum of the Thorlabs APD430C amplified linear mode APD. The experimentally measured integrated noise of this detector (17 nW) is in close agreement with the datasheet (17 nW).

### D. Time Integrating Receiver

The time integrating receiver sums up photocurrent in time. The details of this receiver are in Supplemental Materials V. We model the effects of thermal readout noise the same way as in the above sections but only add noise after each vector-vector product. In experiment we measure the total change in output voltage from the integrator for an integration window by subtracting the voltage at the start of integration from the voltage at readout. This is converted to integrated charge by multiplying by the calibrated capacitance. We then convert from electrical charge to optical energy using electron charge and photon energy. This energy is then divided by the number of MACs performed per integration window, which in the main text is $M = 100$.

Shown in Supplementary Figure 8 is a description of readout noise from the time-integrating receiver used in the main text. Supplementary Figure 8(a) shows the measured values from repeated integration, with the integration time varied between 10 us and 100 us. Figure 8(b) shows the mean integrated signal as a function of integration time. Figure 8(c) shows the rms voltage noise measured from the integrator as a function of integration time. The value of $\approx 220 \mu V$ matches the specifications from the manufacturer. This value gives us our readout noise floor. For the experiments done in the main text we can estimate a per MAC noise floor of $\frac{V_{rms} C_{int}}{M} = 22 aC$. This number is larger than one might expect for a $C_{int} = 10 pF$ integrator, where the readout voltage noise should be $\approx 20 \mu V$. This off the shelf integrator is limited by an internal technical noise with measured noise matching the specification sheet.

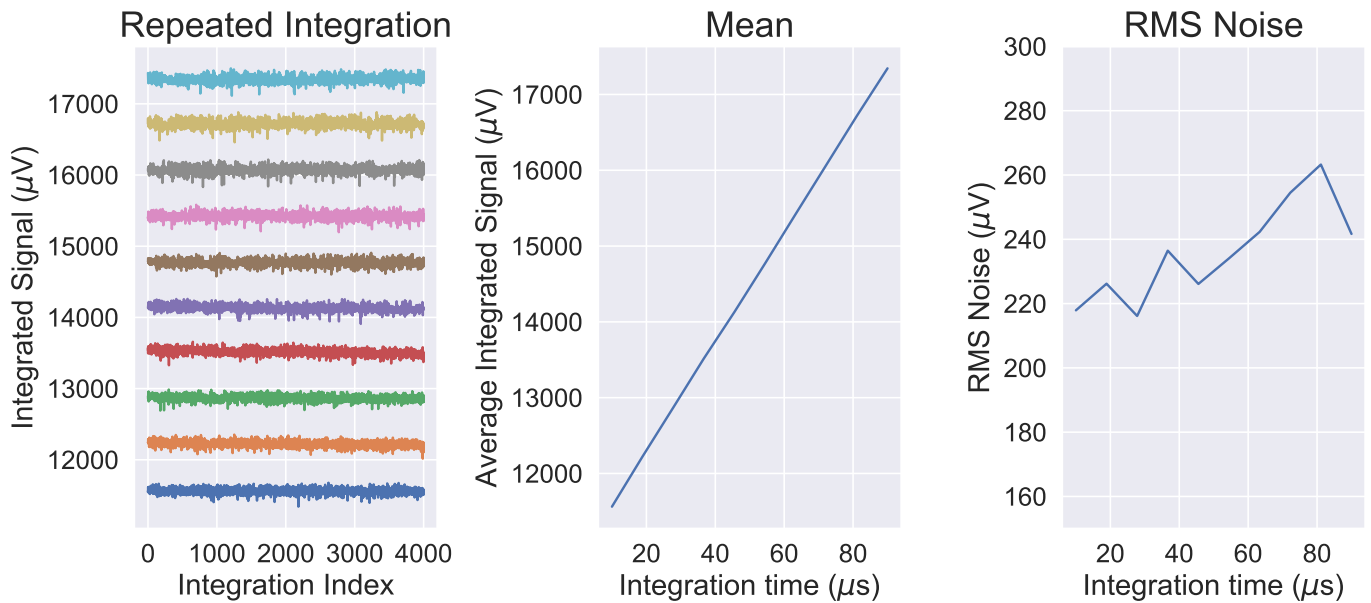

Supplementary Figure 8: Measurement of custom time integrator readout noise. (a) The integrator is measured repeatedly (4000 times) for a fixed optical input power (12 nW) and for different integration times between 10 us and 100 us. This gives rise to 10 timetrace waveforms, each corresponding to a different integration time. (b) The mean integrated value from each timetrace is plotted, showing the integrator is indeed linear in integration time. (c) The RMS readout noise voltage is plotted, showing $\approx$220 $\mu$V of readout noise, matching the specifications from the manufacturer.

### E. Superconducting Nanowire Single Photon Detectors

To find the number of photons used we first perform a calibration of the single photon detector which is detailed in Supplementary Materials XVII. This calibration gives us both a voltage offset and difference in voltage between the measured number of photons. For each readout from the integrator in the SNSPD experiment we subtract the offset so that zero-volts maps to zero-photons and divide by the difference in voltage to get the number of photons per readout. In our experiment we only do one MAC per readout, since for a purely shot-noise limited receiver, such as an SNSPD, time-integration no longer lowers the optical energy per MAC. This can be understood by realizing that independent Poisson random variables add together such that $P(\lambda_1) + P(\lambda_2) = P(\lambda_1 + \lambda_2)$, so time-integrating a shot-noise limited source will have the same SNR as adding distinct measurements together on a computer.

For the simulation plot for the single photon detector we add Poisson noise to each multiply in the model from above, sweeping the number of photons as we do. The Poisson noise is approximated as:

$$P_{shot}(q) = \frac{e^{n_p}(n_p)^q}{q!} \qquad (2)$$

where $n_p$ is the mean number of photons incident on the detector per integration window and $q$ is the measured number of photons in an integration window. We sample from this distribution where $n_p$ is defined by the signal to noise ratio.

## V. TIME INTEGRATING RECEIVERS

In this paper we make use of time-integrating receivers which accumulate charge in time. We use these receivers in the text to sum photocurrent (Figure 3, Figure 4) or sum voltage (Figure 5). For the work done in the Figure 3 and Figure 4 the circuit in Supplementary Figure 9(a) is used which consists of an IR photodiode (FGA01FC) connected to an analog integrating IC (IVC102 from Texas Instruments). In current integrating mode the operational amplifier uses negative feedback to force the inverting input (-) to ground. Current generated by a photodetector generates a voltage across the integrating capacitor by depositing charge. This voltage is readout by the digitizer after integration. The analog switch discharges the capacitor for subsequent integration periods. This IC is configured with it's default 10pF integrating capacitance. The integrator works by producing an output voltage from a generated photocurrent $I(t)$ of:

$$V_{out} = \int dt \frac{I(t)}{C} \propto \sum_i W_i X_i \tag{3}$$

where $W_i$ and $X_i$ are a weight and input vector.

For voltage integration the same IC can be used, but a resistor is placed between the input voltage and input terminal of the integrator. The input voltage is dropped across the resistor generating a current for integration, as is shown in Supplementary Figure 9(b). This generates an output voltage of:

$$V_{out} = -\int dt \frac{V_{in}(t)}{RC} \propto \sum_i W_i X_i \tag{4}$$

where the negative sign is a convention of the current direction. When operated as a voltage integrator we flip the sign of the measured voltage in software to avoid confusion, though this does not change operation or calibration of the system.

Supplementary Figure 9(c) shows an example waveform when using the integrator. When open the switch allows charge to accumulate and when closed the switch resets the charge on the integrating capacitor. Because of the finite on and off impedance of the analog switch two effects can be observed:

1. An offset voltage can be seen when the circuit is being reset. There are two sources of this voltage: offset voltages of the operational amplifier used and the finite resistance of the switch. For the IVC102 circuit used in our experiments this offset value is $\approx 500\mu$V. The finite resistance of the switch makes the circuit look like a transimpedance amplifier when closed. This value for the IVC102 is 1500 Ohms. For experimental results in the main text this value does not become a significant contribution, since we typically operate with nanowatts of light.

2. After the switch opens charge is injected on the integrating capacitor. This charge originates from the charge stored on the gates of the analog switch. Several useful references on switched capacitor/integrator circuits are cited [10, 11]. This source of offset is deterministic and removed by calibration.

## VI. CALIBRATION OF TIME INTEGRATORS

To enable accurate measurement of the optical energy per MAC of the time-integrating receivers we calibrate each receiver. We send a known amount of light into each receiver and integrate for a fixed time. We then measure the readout voltage using a digitizer. All integrators agree well with the manufactured integration capacitance of 10pF. Example integration waveforms from the 16 integrators in the setup are shown in Supplementary Figure 10.

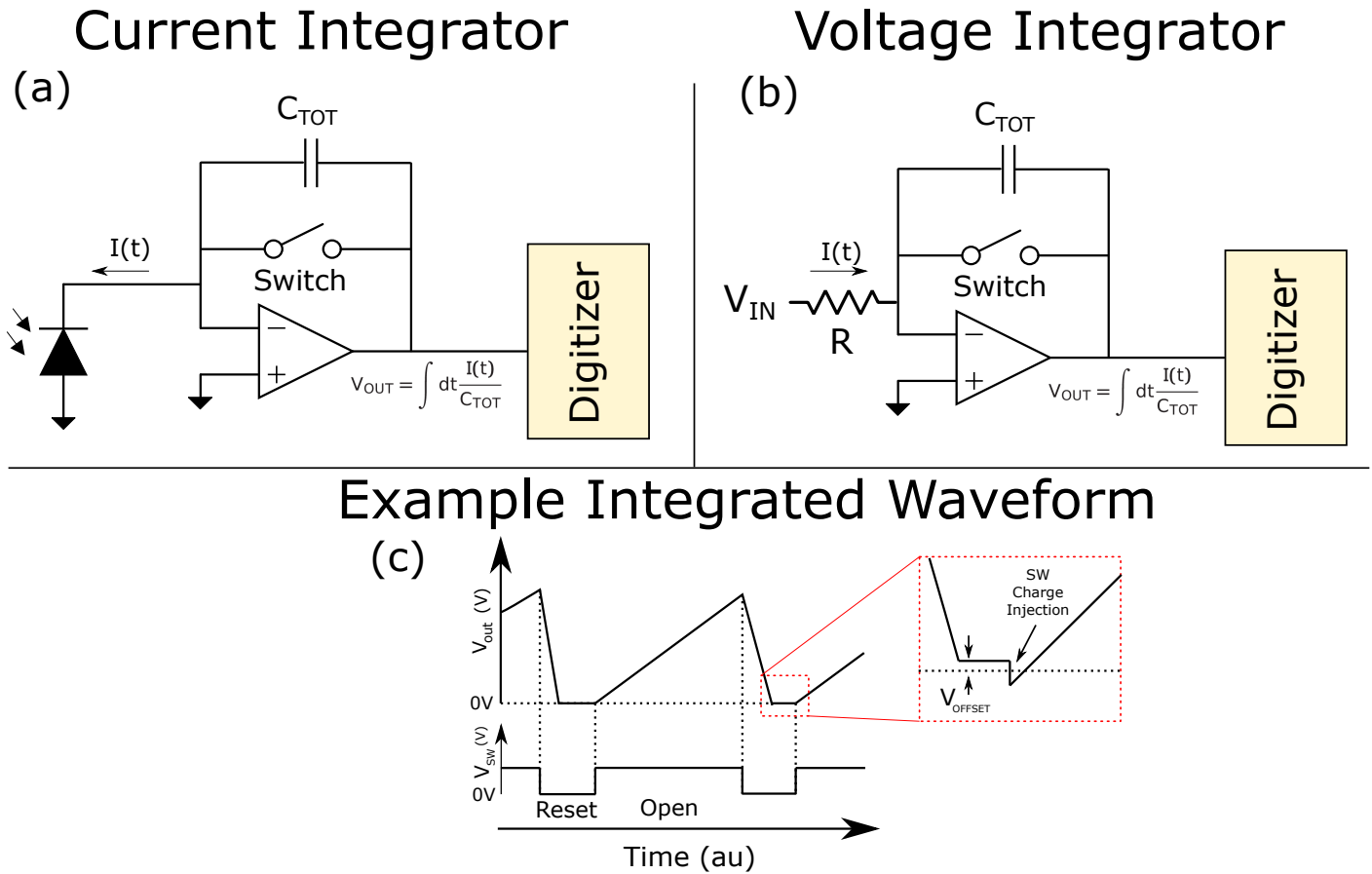

Supplementary Figure 9: Explanation of time-integrating receivers. (a) Current integrators consist of an operational amplifier, integrating capacitor, and analog switch. (b) To integrate voltage a resistor is integrated across the input to the integrator, generating a current. (c) An example integration waveform showing a linear ramp in output voltage for a fixed input optical power.

## VII. TIME INTEGRATING RECEIVER BANDWIDTH

One property that differentiates time-integrating receivers from amplified receivers is the bandwidths available for accurate computation. While the bandwidth of amplified receivers is limited by either the RC time constant of its amplifier or carrier transit times in the photodiode, the fundamental bandwidth limit for a time-integrating receiver is the absorption spectra of the photodiode (10's of THz).

Above the bandwidth of amplified photoreceivers the waveform becomes distorted and computing accuracy drops [12]. Researchers can use methods such as waveform pre-emphasis to compensate for the finite bandwidth of a computing channel, but are ultimately hamped by finite modulator or photodetector bandwidths. As a reference, modern germanium photodiodes in a silicon photonics process have quantum efficiencies of $\approx 1$ [13] with bandwidths of 40GHz readily available in commercial CMOS foundries. Detectors with bandwidths approaching 300GHz, where the intrinsic region of the photodetector is significantly decreased to enable fast carrier drift times, have recently been demonstrated in commercial foundries processes [14].

Time integrating receivers, fundamentally, count photons. Because of this, so long as a photon is absorbed by an absorption medium the generated electron-hole pairs are accumulated on the integrating capacitor. As a result, if a theoretical 1THz optical modulator existed which could encode input and weight waveforms, a time-integrator clocked at 1MHz could readout the $10^6$ MACs after accumulation. Supplementary Figure 11 is an experimental demonstration of this, where optical pulses 1ns wide are sent at a time integrator, clocked with a 10us integration window. As the number of optical pulses inside of the integration window increases the measured output signal increases linearly, corresponding to photon counting.

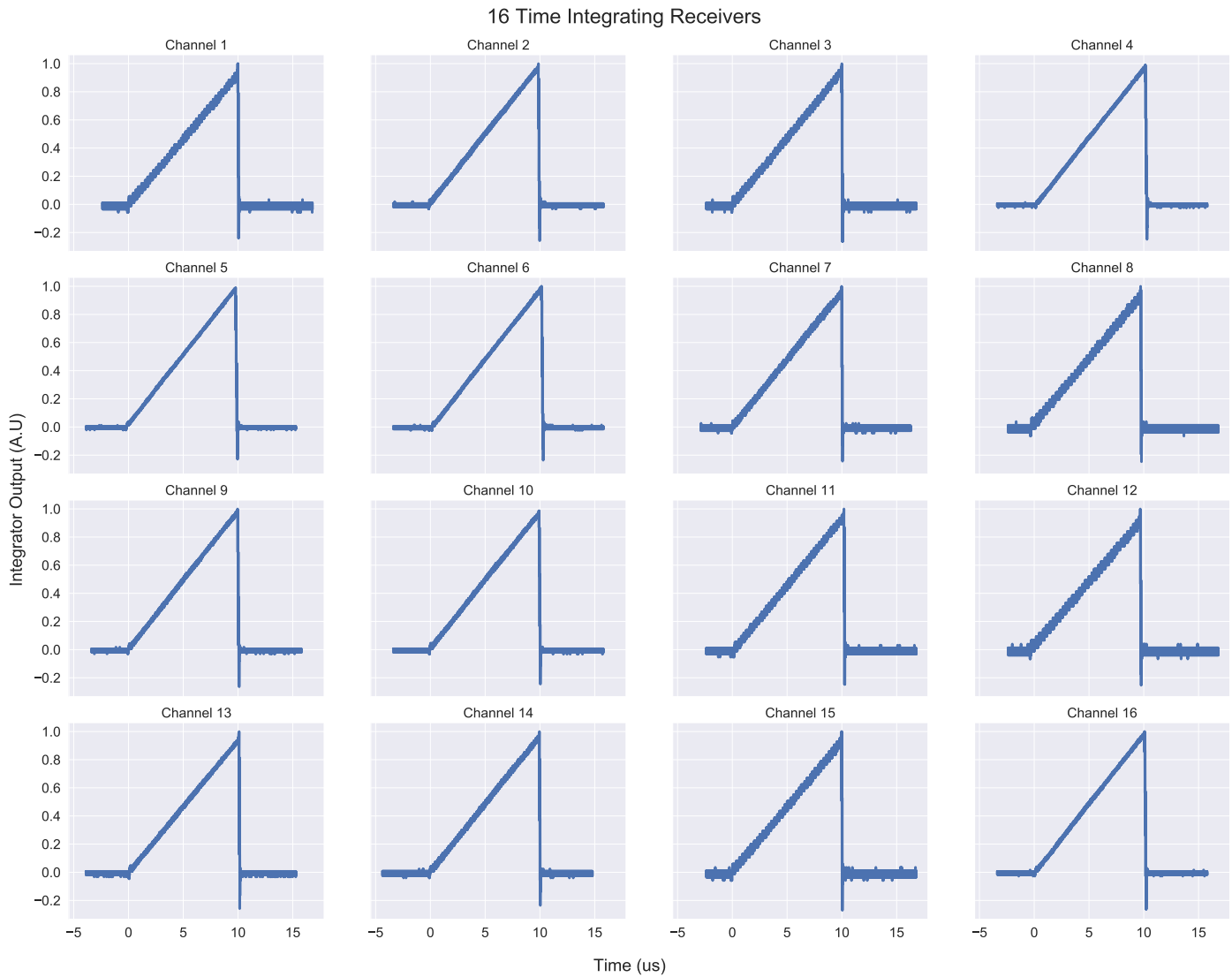

Supplementary Figure 10: Integration waveforms from 16 time integrators. A fixed amount of optical power enters each integrator and integrated for 10us before discharging. Noise on top of each waveform is a result of digitizer resolution settings.

## VIII. EYE DIAGRAM MEASUREMENT

To generate the eye diagram for the lithium niobate modulator seen in Figure 2 in the main text we made use of a 25GHz arbitrary waveform generator (Tektronix AWG70002), 12GHz photodetector (Newport 1544-A) and 10GHz/40GSPS oscilloscope (DSO81004A). Because of the limited bandwidth of both the photodetector and the oscilloscope and the high cost of high bandwidth test and measurement equipment we are only able to characterize modulators up to 10GHz of bandwidth. To generate this eye diagram we used standard pseudo-random bit sequences generated from matlab.

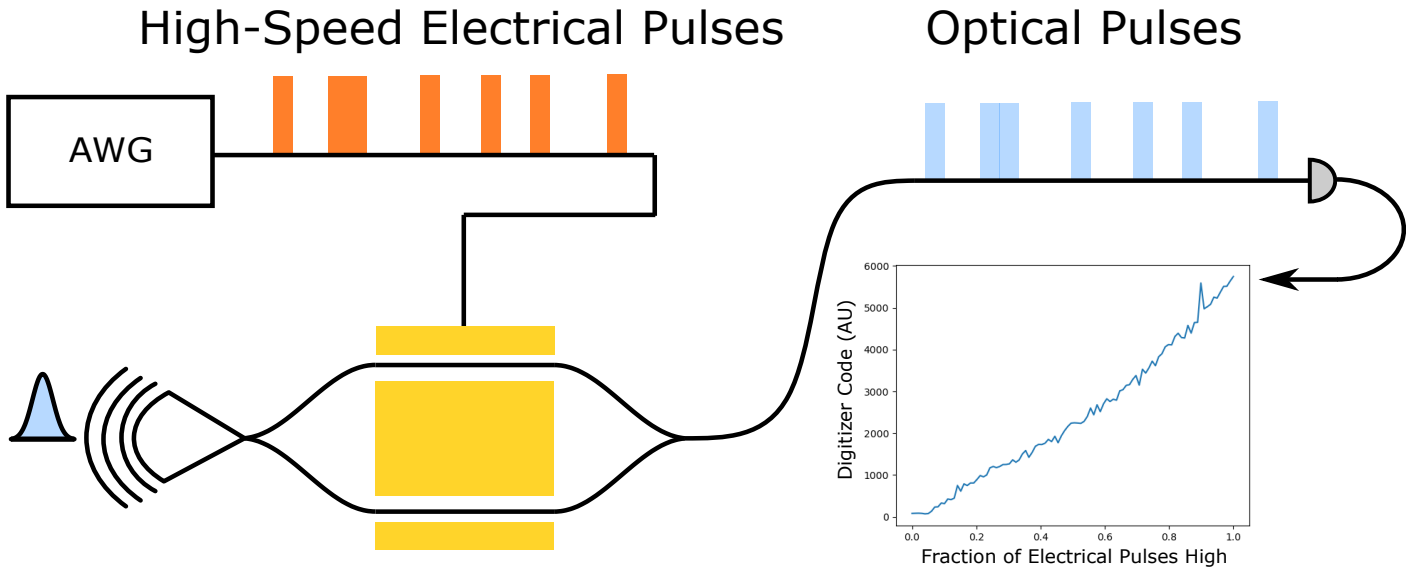

Supplementary Figure 11: Time integrating receiver bandwidth measurement. The on-chip silicon modulators are biased to the extinction point and driven by a train of 1ns electrical voltage pulses. We sweep the number of electrical pulses which are high in an integration window and see a linear increase in the measured voltage after the integrator.

The 50GHz eye diagram of the silicon modulators was measured using a Agilent 86100C Digital Communications Analyzer, the details of this measurement can be found in [15].

## IX. EFFECT OF SYSTEM LOSSES

System level losses demonstrated in this paper can reach up to 30dB per wavelength, primarily limited by the losses coupling light in and out of the chip and combining wavelengths together. In modern silicon photonic processes these coupling losses can reach <1dB using either edge couplers [16] or grating couplers [17] with 100nm and 80nm of optical bandwidth respectively. Further improvements can be achieved with non-standard technologies such as integrating a metalic mirror below a grating coupler which can achieve 85% coupling efficiency [18], making use of heterogeneous integration through angled couplers [19], or free-form 3d printed optical couplers [20]. To understand realistic datarates which are possible using the time-integrating receivers demonstrated in this paper we perform an example calculation assuming existing telecom devices. Assuming a starting laser power of 10mW (10dBm), light enters a weight server where it experiences 10dB fiber-to-fiber loss (4dB from coupling, 3dB from modulator insertion loss and 3dB from loss on passive components). Then, using 70km of deployed fiber with 0.14dB of loss per kilometer (a typical value in the optical C and L bands) 10dB of loss is experienced. At the receiver another 6dB of loss is experienced (1.5dB for coupling, 3 dB for the input activation modulator, and 1.5dB for other passive components) before being absorbed by a photodiode. In this conservative estimate we have 25uW of light at the client time-integrating detectors (-16dBm). Assuming 100aJ/MAC of optical energy is required at the receiver then the receiver can compute at 250GHz for each wavelength. As we discussed above in Supplementary Materials VII the bandwidth of time-integrators is not limited by RC time or carrier transit time, but by the absorption spectra of the photodiode.

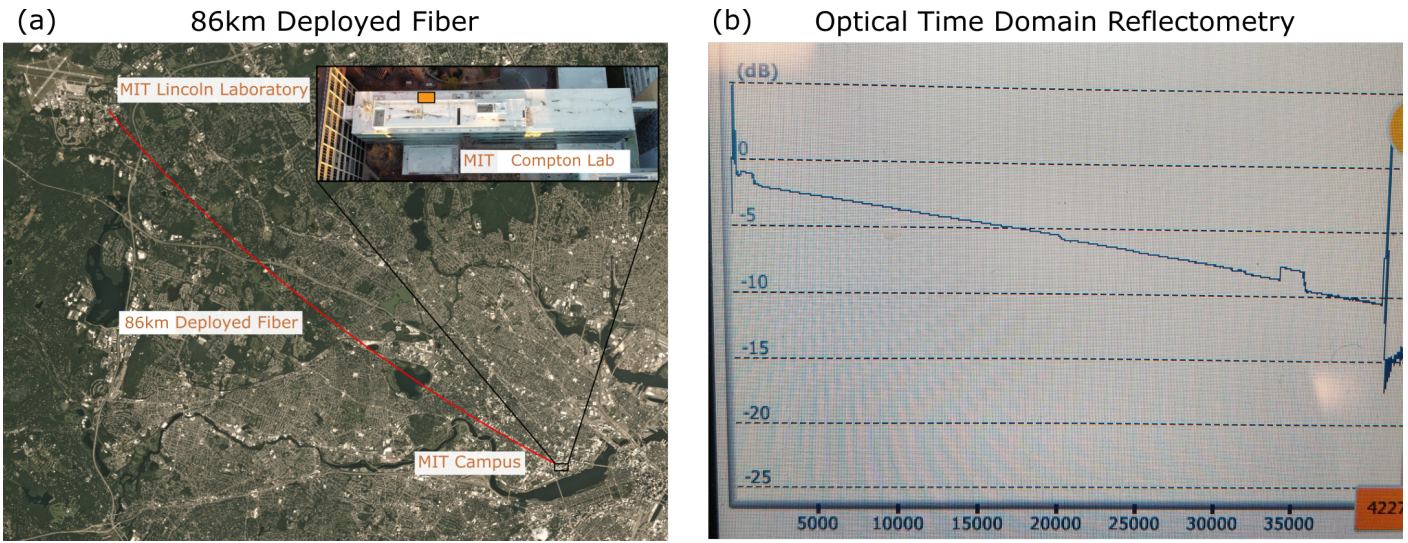

Supplementary Figure 12: (a) Satellite imagery showing the greater Boston area and an estimate of the path taken by the fiber between MIT and MIT Lincoln Laboratory. (b) Screenshot of an optical time domain reflectometry (OTDR) measurement of one strand of the fiber taken at 1550nm.

## X. DEPLOYED FIBER LINK

The fiber connecting MIT and MIT Lincoln Laboratory is composed of two 43 km long fiber strands, each following the same physical path. Because of constraints on physical access to Lincoln Laboratory, an air force research facility, the fiber is placed in 'loopback', sending light along the first strand to Lincoln Laboratory and back along the second strand. While not ideal, this still presents an exciting testbed for edge computing systems. The lincoln fiber is a real-world fiber with drifting polarization, high loss, dispersion and reflection. In addition, we operate in a regime where the weight server and client have their signals time-delayed from each other by $\approx 480$ us, shorter than the operating bandwidth along the fiber. As an example of the problems associated with using a deployed fiber, Supplementary Figure 12(b) shows an Optical Time Domain Reflectometry (OTDR) trace taken along one strand of the fiber. OTDR is a technique where a pulse of light is sent into a fiber or optical component and a time-trace is measured of the reflected signal. This reflection, which can originate from loss due to scattering in the fiber or fiber discontinuities, can be plotted as a function of distance to inform the user about the "health" of a fiber. The decreasing line in Figure 12(b) shows propagation loss and 'jumps' in the reflectometry signal show discontinuities in the fiber. These jumps, which could originate from fiber-to-fiber connections or splices from points where the fiber has broken previously, reflect signals. Reflected signals in an optical fiber or waveguide can lead to Fabry-Perot effects that may lead to intensity instability in larger fiber systems. The loss along each strand of fiber is $\approx 22$ dB, primarily limited by defects such as splice quality.

For the experimental demonstration in the paper, showing 98.8% accurate classification across 3 THz on the deployed fiber, we made use of two lasers parked at 191.6THz (ITU grid channel 16) and 194.6THz (ITU grid channel 46). Data was encoded onto these two channels by "chunking" the problem up along the wavelength dimension, as described in Supplmentary Material 16. Because of loss along the fiber the system is operated at a speed of 1kHz. We would like to stress this demonstration is proof-of-principle, we have shown that at least 3 THz of bandwidth is available for simultaneous classification at the client. A company seeking to make a commercial product will be able to fill this bandwidth using either a bank of lasers or a comb source and a bank of modulators/filters.

## XI.  COMPENSATING LOSSES USING AN ERBIUM DOPED FIBER AMPLIFIER

Erbium doped fiber amplifiers (EDFAs) are routinely used in datacenters and long-haul optical communication links to boost a weak signal as it experiences loss through a fiber. One downside of EDFAs is additional noise that they add to the amplified signal, limiting the total amount of amplification and bandwidth possible in a communication link. In Netcast an EDFA can be used to take data from the smart transceiver and amplify it before reaching the client, increasing the potential bandwidth possible for a fixed detector optical sensitivity.

An analytical derivation of the input-output relationship of an EDFA can be found in Yariv in the section on EDFAs [21]. The important derivation from this section is that amplified spontaneous emission power scales as:

$$\mathrm{P_{ASE}} = \mu h\nu \Delta \mathrm{f_{opt}} (G - 1) \tag{5}$$

where $\mu$ is the population inversion factor (the ratio of population in the excited state to the difference of populations in excited and ground states, typically $\approx 1$ for gains larger than 1), $h\nu$ is the photon energy, $\Delta f_{opt}$ is the optical bandwidth of the wavelength channel, and $G$ is the gain.

Assuming we want to compensate for realistic chip coupling losses, modulator losses, wavelength multiplexer losses, and fiber losses, which could amount to $\approx 20\mathrm{dB}$, a gain of 100 will be needed. For a standard 100GHz optical channel, similar to what is found in communication systems today, the ASE noise at 1550nm will be $\approx 1\mu\mathrm{W}$ per channel. Assuming all of the 100GHz bandwidth is used for computation, this amounts to 10aJ/MAC of ASE noise at the receiver. This number can be improved by using narrower bandwidth filters, such as resonant ring or disk filters that match the used transmitting bandwidth or using lower gain.

To understand EDFAs further we measure the input output characteristics of a commercial EDFA. Shown in Supplementary Figure 13 is experimentally measured data for the performance of a commercial EDFA (Oprel OFA17D-1221M). A commercial 100GHz wide fiber filter (Fiberdyne DWDM filter) is used to remove amplified spontaneous emission (ASE) noise. The pump power of an internal $\approx 980$nm pump laser, given by the EDFA as a percentage of the maximum pump power, is increased while input power is swept. We observe that for low pump powers the EDFA acts as an attenuator, adding a low amount of ASE noise. At  10-15% pump power the EDFA achieves gain of the input signal. At high pump powers and high input powers saturation occurs, when the pump is depleted from conversion to both the amplified input signal and ASE noise.

As an example of the gain possible from using an EDFA to measure loss through the smart transceiver Figure 2(b) from the main text is generated by taking the output of the chip and passing it through an EDFA to boost each weight signal.

## XII.  NETCAST USING COHERENT DETECTION

Recent work has demonstrated that optical coherent detection can natively enable computing when the signal and local oscillator encode inputs and weights in their field amplitudes [22]. Conceptually, coherent detection is the mixing of two signals, a weak signal field and a strong local oscillator field. In the radio-frequency domain this mixing is done on electrical mixers while in the optical domain mixers such as beamsplitters and directional couplers are used.

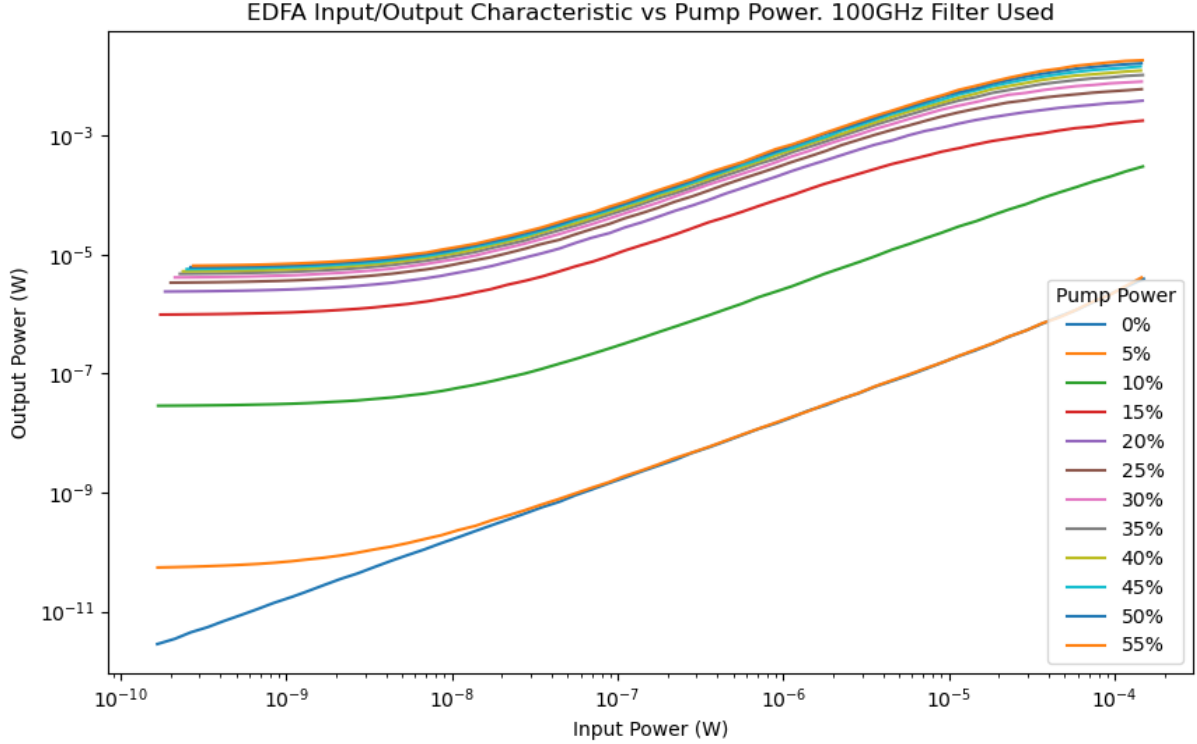

Supplementary Figure 13: Experimentally measured EDFA input output characteristic for a single 100GHz optical channel.

When coherent detection is performed the generated electrical current at the receiver is given by:

$$I_{coherent} \propto E_{sig} E_{LO} \propto \sum_i X_i W_i \tag{6}$$

showing that if $E_{sig} < E_{LO}$ then a weak signal field can be amplified at the receiver by a strong local oscillator. This enables a weak signal to more easily reach the shot noise limit. The same principle applies to Netcast, and is illustrated in Supplementary Figure 14. In this scheme, a comb of local oscillators can mix with a received comb of frequencies encoding the weights. The combs have their free-spectral-ranges and resonant frequencies locked using integrated tuners, such as thermal heaters in silicon nitride. These locked combs now only require a single wavelength to be phase stabilized for all wavelengths to be phase stable, removing significant complex digital signal processing which is found in existing coherent transceivers [23–25]. Emerging frequency comb technologies include combs made in thick silicon nitride and electro-optic combs made in thin-film lithium niobate [26–28].

## XIII. RELATIVE INTENSITY NOISE

In the main text we discussed how thermal noise, detector shot noise and laser relative intensity noise (RIN) can be fundamental limiting factors for the performance of the system. While thermal and shot noise were addressed in depth in the main text, laser relative intensity noise will be addressed here. Laser's are not perfect oscillators and experience decoherence in the output laser field on $\approx \mu$s time scales [21]. This decoherence leads to laser phase noise, linewidth, and intensity noise. Fundamentally, the relative intensity noise from a laser can approach a shot-noise/quantum limit, given by:

$$\text{RIN} = \frac{2\text{h}\nu}{\text{P}_{\text{av}}} \tag{7}$$

where $h\nu$ is the photon energy and $P_{av}$ is the average output power of the laser. This RIN is in units of $\text{Hz}^{-1}$, and the dimensionless quantity is found by multiplying by the optical bandwidth seen by the receiver. For 1550nm semiconductor lasers the shot-noise limited relative intensity noise for a 20mW output power laser is $\approx 1.3 \times 10^{-17} \, \text{Hz}^{-1}$. Conventionally, this number is reported in units of decibels relative to the carrier per unit bandwidth. For a shot-noise limited RIN this is -169dBc/Hz. Modern semiconductor lasers approach this number, achieving $\approx$ -150dBc/Hz as can be seen in Supplementary Figure 15 and with mass produced transceiver tunable lasers achieving <-140dBc/Hz [29]. Assuming -140dBc/Hz of RIN noise and a receiver wavelength channel of 100GHz we find a receiver SNR for mass produced tunable lasers of >30dB (1000), larger than is required for neural network computation [30]. This implies that laser relative intensity noise will not be a significant concern for Netcast.

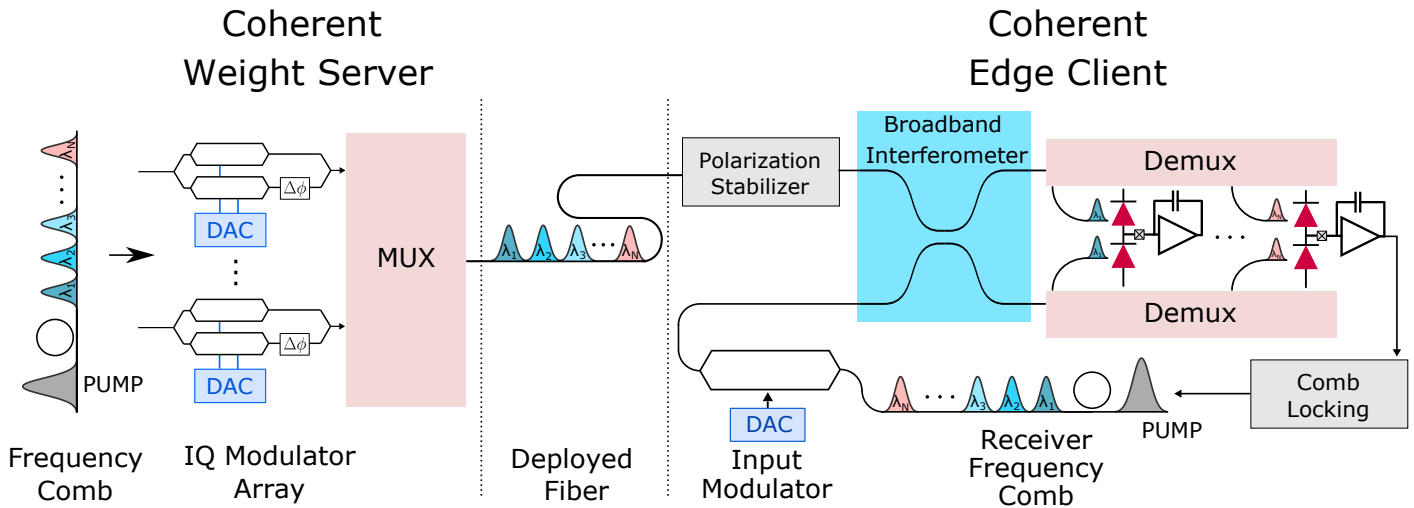

Supplementary Figure 14: Schematic of coherent netcast. An optical frequency comb's output is split using a wavelength multiplexer and each comb tooth is sent to an IQ modulator (which gives both phase and amplitude control). Each comb tooth is then recombined and deployed over optical fiber to the receiver. At the receiver a separate frequency comb is generated which is coherent with the weight server's comb. A broadband modulator encodes input data onto all receiver comb teeth simultaneously. The received weight comb and generated input comb mix in a broadband interferometer, before being split and sent to separate time-integrating homodyne detectors.

## XIV.   MODEL USED FOR IMAGE CLASSIFICATION

The model used in this paper is a 2 hidden layer fully connected model with layer sizes [784 → 100 → 100 → 10]. Biases are included on each layer and are added into the result of matrix-vector computation in post. This model was trained using a neural architecture search algorithm, similar to [31]. Our model was trained using an Nvidia Tesla K40 GPU donated by the Nvidia Corportation which we would like to again thank for their contribution to our research.

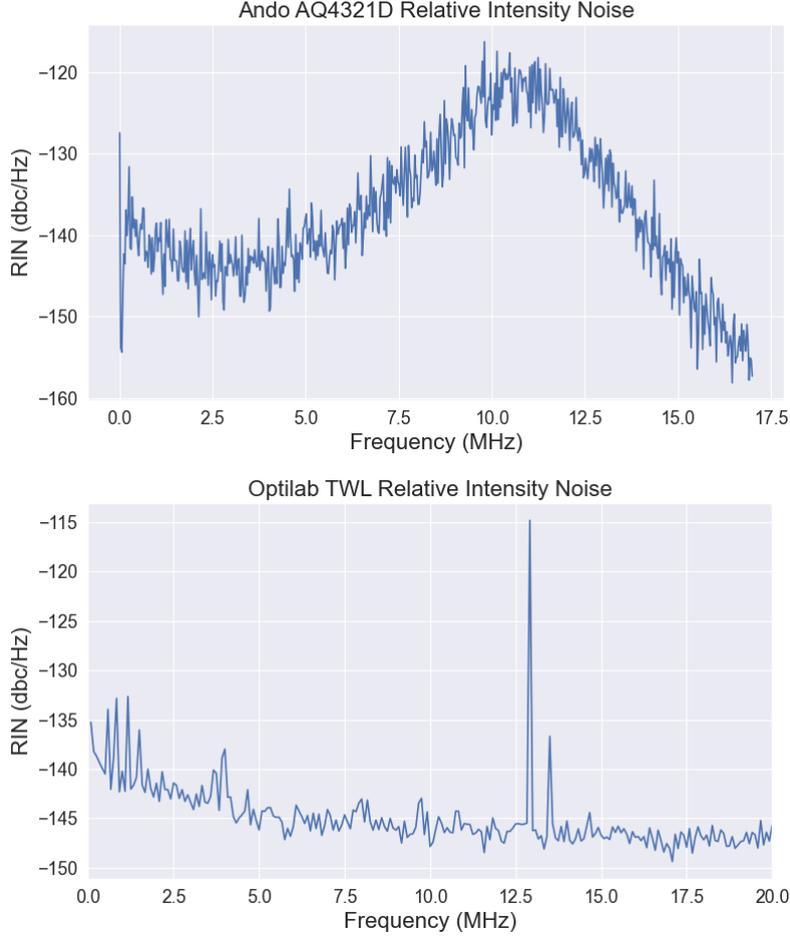

Supplementary Figure 15: Relative intensity noise measured for the two lasers used for experiments: an Ando AQ4321D tunable laser and Optilab TWL tunable laser. The peaks in the TWL spectrum a pilot tones the manufacturer applies to the laser.

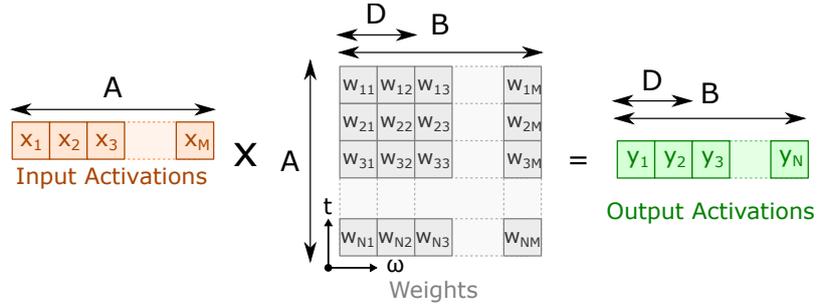

Supplementary Figure 16: Mapping matrix-vector multiplication onto hardware of a finite size. A matrix of size (A,B) is broken into chunks of size (A,D) to map onto hardware with D wavelengths. Over $\lceil B/D \rceil$ steps the full matrix-vector multiplication is performed.

## XV.  MAPPING MATRIX-VECTOR MULTIPLY ONTO HARDWARE

In Netcast our computation is performed in a time frequency basis. Time is theoretically infinitely extensible while maintaining a fixed system clock rate. However, frequency is a finite resource. Here we discuss how a problem of a fixed size can be mapped onto this finite resource so that neural networks of any format can be run on the Netcast system.

Show in Supplementary Figure 16 is an example of a matrix-vector multiplication with a matrix size of $(A, B)$ running on hardware with $D$ wavelengths. Over $\lceil B/D \rceil$ time integration windows the matrix-vector multiplication is performed.

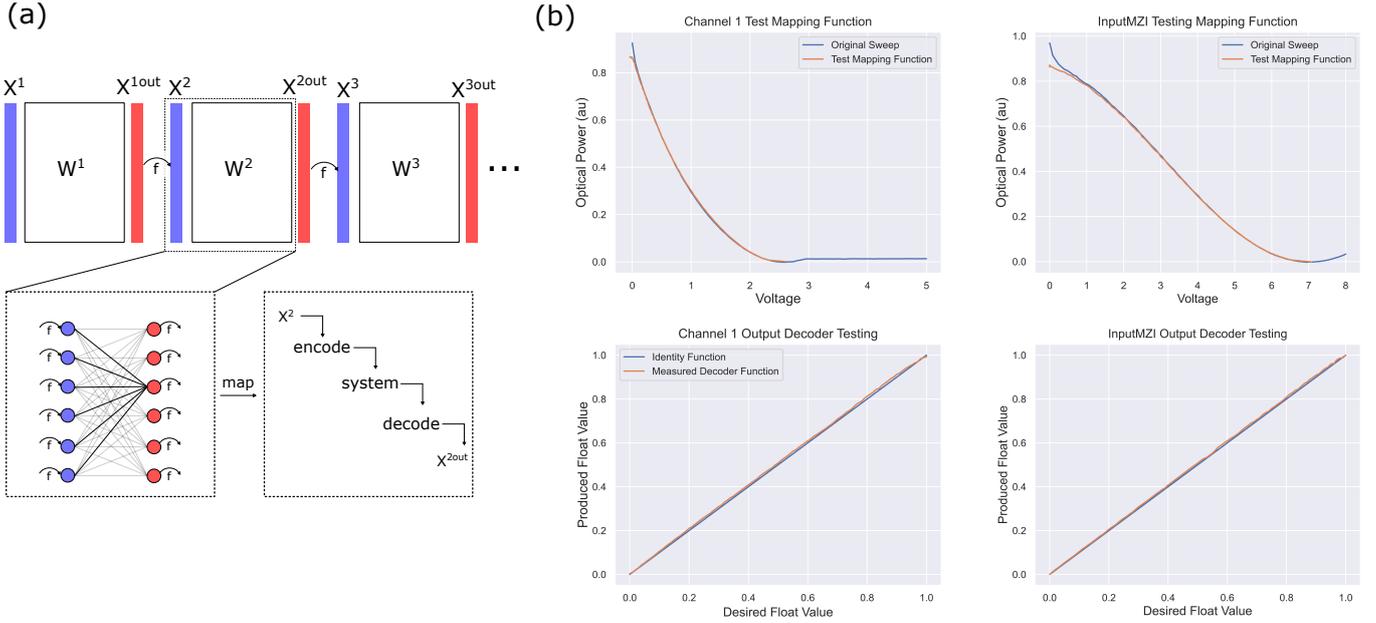

Supplementary Figure 17: Illustration of calibration of Netcast system. (a) description of floating point to optical hardware mapping involving encoding function, system dynamics, and decoding functions. (b) Example calibration data of the optcial hardware showing a voltage sweep of the Silicon modulator (top left) and Lithium Niobate (top right). As well as tests for the decoding function in the bottom left and right.

## XVI.  CALIBRATION OF OPTICAL SYSTEM

In order for optical hardware to compute accurately a repeatable and accurate method of calibration is required. In this paper we use new calibration procedures designed for time-integrating receivers. The general structure of how a problem is mapped onto hardware is shown in Supplementary Figure 17.

### A.  Encoding

First, we start by considering the problem, shown in Supplementary Figure 17(a). We must take a floating point vector-matrix computing problem and creating mappings onto optical hardware that convert the input activations ($X^2$) and weight values ($W^2$) onto voltages on their respective optical modulators using encoding functions and decoding the measured signal back to floating point output activations using a decoding function. To do this, we must first create a mapping from voltage to optical intensity, shown in Supplementary Figure 17(b). We sweep the applied voltage on all weight modulators in the system and the input modulator and measure the transfer function from the respective photodetector at the receiver. Then, we fit a third order polynomial between the maximum and minimum point of the transfer function using numpy's Chebyshev polynomial function [32]. Now, all weight modulators have a function $f_1, f_2, ..., f_N$ which takes as input a desired optical intensity in the range $I_{min}, I_{max}$ and generates the voltage that must be applied to generate this intensity. Next, we need to know how to map from our floating point computing problem onto the optical hardware. We normalize the vector-matrix multiplication to be in the range $(-1, 1)$ and after computing with the setup and multiply by the appropriate scaling factors to re-normalize. Because we only use two optical modulators back to back in our setup we can only encode non-negative values. As a consequence, negative values are encoded on separate timesteps. This is not fundamental and can be overcome using methods described in Supplementary Materials XXIV. We now map the range $[float_{min} = 0, float_{max} = 1]$ linearly to $I_{min}, I_{max}$ using the linear interpolation:

$$I_{out} = \frac{I_{max} - I_{min}}{float_{max} - float_{min}} \cdot float_{in} + I_{max} - \frac{I_{max} - I_{min}}{float_{max} - float_{min}} \cdot float_{max} \tag{8}$$

Then the corresponding channel intensity to voltage mapping function $f_i$ is applied to encode data onto the modulator.

### B. Decoding

To decode data out of the system the photoreceiver needs to understand the measured photosignal. For non time-integrating systems we make use of the fact that $0 \cdot 0 = 0$ and $1 \cdot 1 = 1$. We encode zeros onto all modulators in the system and measure the generated photosignal at each receiver, designating this value as $float_{min}$. We do the same by applying floating point 1 values to all modulators, measuring all photoreceivers, and calling the measured signal $float_{max}$. We now have a linear mapping between measured photosignal at each receiver and floating point numbers. Time integration calibration is performed in the same way, but the decoder for a vector of length $M$ uses the range $[float_{min} = 0, float_{max} = M^2]$. The non time-integrated decoder functions are tested on the bottom left and right of Supplementary Figure 17(b) where the Lithium Niobate and Silicon modulators, respectively, are held to floating point 1 and the other modulator is linearly swept from floating point 0 to 1 at the encoder while the decoder decodes the value. If the encoder and decoder are both well calibrated the resulting line should be the identity function, corresponding to perfect calibration of encoders and decoders for that modulator. The result of this calibration procedure are seen in the main text in Figure 3(b) where two arrays of 10,000 floating point numbers are multiplied, giving a uniformly distributed computing result with $\approx 8$ bits of accuracy.

## XVII. SUPERCONDUCTING NANOWIRE SINGLE PHOTON DETECTOR CALIBRATION

In order to create Figure 5(c) and Figure 5(d) in the main text we need a way of mapping from the measured voltage at the output of the integrator to the number of photons within an integration window. To do this, we send a fixed amount of optical power into the single photon detector and measure the generated voltage from our integrating circuit 10,000 times. This generates the histogram shown in Supplementary Figure 18. Distinct voltage levels are formed with equal spacing, corresponding to distinct numbers of photons. From this, we find a linear mapping between the peaks in this histogram and the number of photons which is received. This mapping is used for both the Poisson statistics plot shown in Figure 5(c) and calculating the number of photons per MAC in Figure 5(d). For our experimental demonstrations using the single photon detectors we clock the system at 30kHz, allowing us to operate away from the dark counts of the detectors (which were measured to be <1000 per second, implying a dark count probability of <3% per integration window). Our maximum operating speed is set by saturation power on the single photon detector, which is $\approx 100$fW or $\approx 10^6$ photons per second. Above this power the SNSPD latches and must be reset.

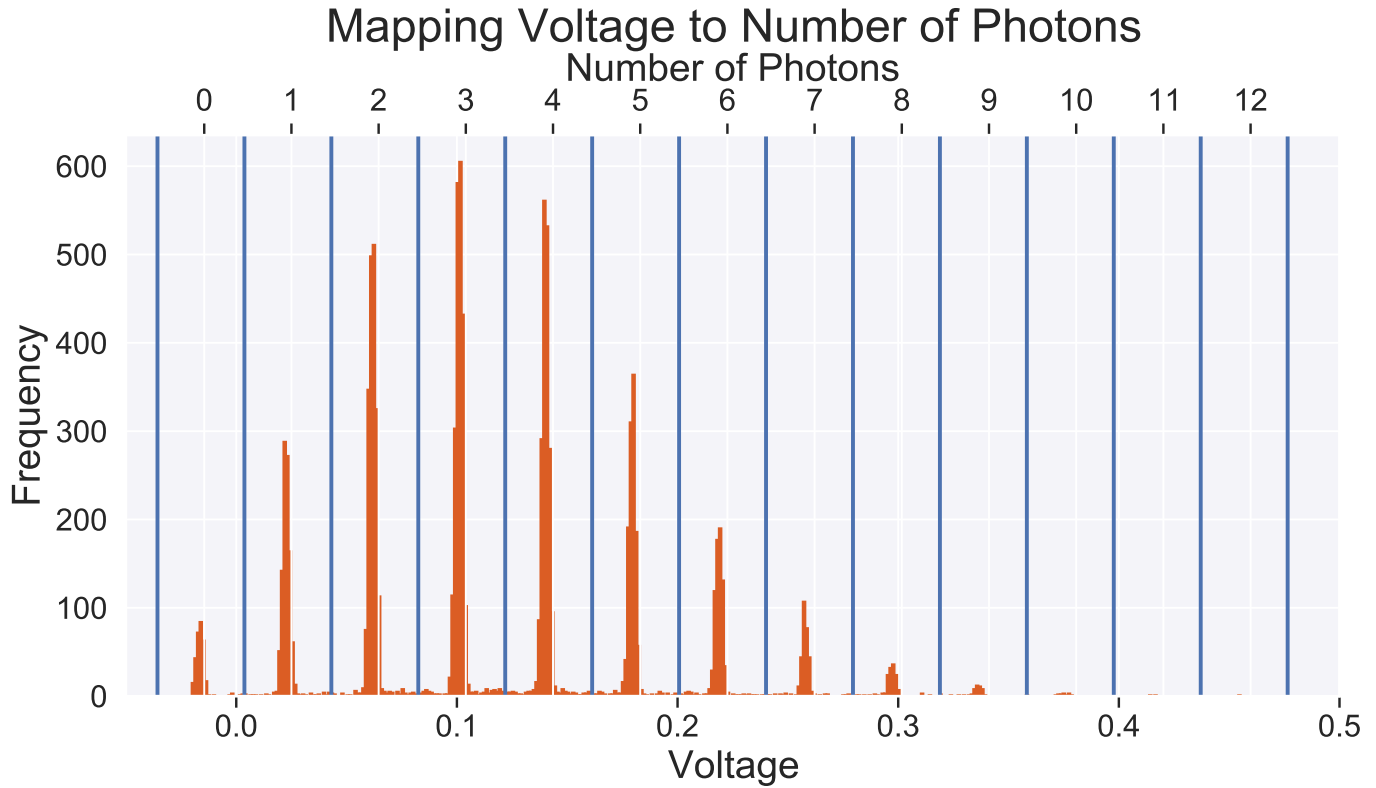

Supplementary Figure 18: Mapping between voltage generated from receiver circuitry and number of photons from single photon detector. The bottom x-axis shows the measured voltage at the receiver. Blue lines here delineate boundries used in Figure 5(c) to generate Poisson statistics. The top x-axis enumerates the photon number of each bin.

## XVIII. LESS THAN ONE PHOTON PER MAC EXPLAINED

In the main text, a key result is that the Netcast system can scale to be shot noise limited, achieving less than a single photon per MAC with high accuracy. This result may seem unintuitive at first: what does it mean to have less than one photon on each multiplication step? How can a model achieve high accuracy with less than one photon on each step? To help the reader understand this we created Supplementary Figure 19, which details how during vector-vector multiplication each multiplication step of the vectors can have less than one photon, but the overall result can have more than one photon and an SNR greater than 1.

## XIX. ENERGY COMPARISON OF STANDARD PHOTODIODES, LINEAR MODE APDS AND GIEGER MODE APDS

To enable computing in photon starved environments we must make use of the best photodetectors possible for Netcast. While time-integration with standard photodiodes gives good results, further improvements are possible by making use of photodiodes with intrinsic gain, such as avalanche photodetectors (APDs). However, the gain of avalanche detectors comes at the cost of increased electrical energy consumption at the receiver, required because of the large bias voltages that must be applied to APDs to bring their material close to breakdown.

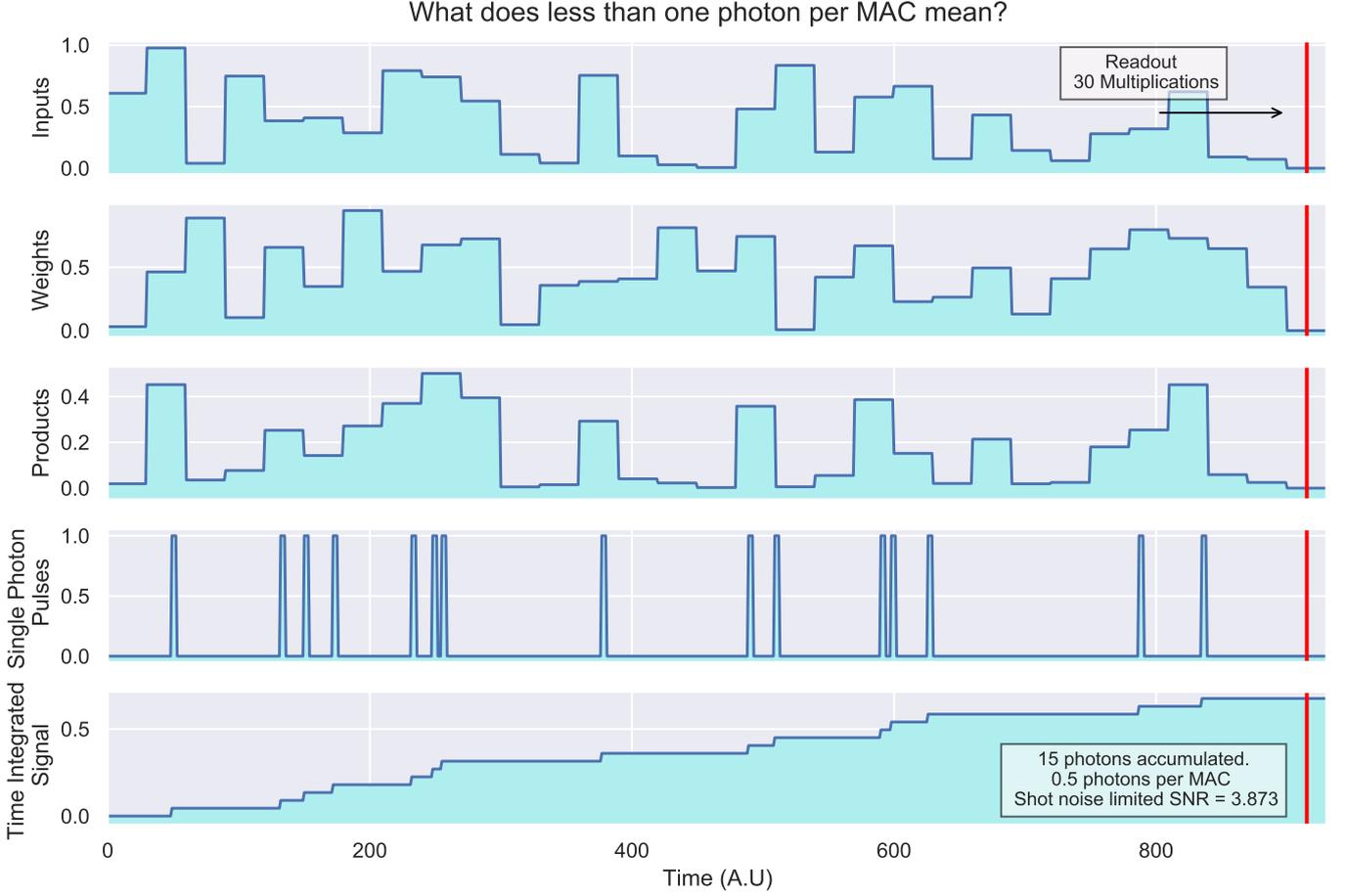

Supplementary Figure 19: Example simulation to illustrate the concept of less than one photon per MAC. A stream of input and weight floating point values are applied to modulators, attenuating a laser. High output product values have a higher probability of generating a single photon pulse. The output signal is the time-integral of the pulses. At the output of this vector-vector product with 30 MACs in it, 15 photons have been accumulated. This results an SNR of $\approx 3.9$, while each MAC on average had 0.5 photons in it.

The energy that a photodetector must pull from a power supply can be calculated as:

$$E_{det} = Q_{det} V_{bias} = \frac{q \eta E_{mac}}{h \nu} V_{bias} \tag{9}$$

where $E_{det}$ is the electrical energy pulled from the power supply per MAC, $Q_{det}$ is the charge pulled per MAC, $V_{bias}$ is the bias voltage applied to the photodiode, $q$ is the electron charge, $\eta$ is the quantum efficiency of the photodiode, $E_{mac}$ is the optical energy per MAC of the system and $h\nu$ is the photon energy. For the time-integrating receiver used in the main-text this energy consumption is minimal, as the bias voltage across the photodiode is the input offset voltage of the integrator, which is mV scale. For a standard Germanium photodiode, operating with $V_{bias} = 1$ and using a transimpedance amplifier at $E_{mac} \approx 10^{-15}$ J of energy per MAC we find $E_{det} \approx 1$ fJ/MAC, which could become a limitation of the receiver energy consumption when emerging technologies for modulators and data conversion are incorporated.

Linear mode and Gieger mode APDs modify Supplementary Equation 9 by the intrinsic gain of the APD (also referred to as the "M" factor, not to be confused with the M value used in the main text which refers to time-integration length):

$$E_{det} = Q_{det}V_{bias} = \frac{q\eta M}{h\nu}\frac{E_{mac}}{M}V_{bias} = \frac{q\eta E_{mac}}{h\nu}V_{bias} \tag{10}$$

where the amount of charge which is generated at the APD is increased by the intrinsic gain, but the optical energy per MAC required decreased by a factor of M, assuming a thermal-noise limited operating regime. While at face value it may seem that the energy consumption of the APD has not changed, the bias voltage of an APD is significantly larger than a standard photodiode to reach breakdown. For our consideration, we will use consider Silicon and Indium Gallium Arsinide (InGaAs) as example materials used as the multiplication region of an APD. Both Silicon and InGaAs have breakdown voltages of $\approx 40V/\mu m$ [33, 34]. Assuming a waveguide coupled photodetector in a P-I-N configuration with a $1\mu m$ wide depletion region we find that 40V must be applied to achieve breakdown. This would result in a 40x increase in the amount of energy consumed by the APD. Below breakdown the device operates in the linear-mode regime and at/close to breakdown it operates in Gieger mode. This additional energy consumption could be prohibitive, bringing receiver energy consumption up to $\approx$ 1pJ/MAC, no better than using existing CMOS ASICs. However, in certain applications operating with low photon numbers but higher total energy consumptions may be advantagous, such as when using <1 photon per MAC for secure neural network computation over a local network.

The above derivation and arguement assumes operation in a thermal-noise limited regime, where increasing the detector intrinsic gain enables lower optical energy per MAC. This improvement stops once the shot noise limit is reached. Increasing the gain beyond the shot-noise limit only results in excess power consumption at the receiver, scaling as:

$$E_{det} = Q_{det}V_{bias} = \frac{q\eta M E_{mac-shot}}{h\nu}V_{bias} \tag{11}$$

## XX. EFFECT OF DISPERSION AND NONLINEARITY ON NETCAST

An obstacle to scaling Netcast are the different arrival times of wavelengths at the receiver caused by dispersion in optical fiber. For a single smart transceiver and single client the effects of dispersion can be removed by making use of programmable delays on the transmitter side. However, if multiple clients have weight data deployed to them simultaneously with different lengths of optical fiber then this technique can not be used. For this case, we consider the effects of dispersion on the accuracy of neural network classification. Dispersion presents itself as a form of temporal symbol crosstalk. The effects fo temporal crosstalk are a decrease in the channel capacity, which can be modeled as:

$$C = \frac{2\pi\sqrt{2\chi}}{\log\left(\frac{1}{\chi}\right)}B_{opt} \tag{12}$$

where $C$ is the channel capacity in weights per second, $\chi$ is a temporal crosstalk factor, $B_{opt}$ is the optical bandwidth [35, 36]. The temporal crosstalk factor is represented as:

$$\chi = \frac{\text{Maximum Delay Time}}{T} = \frac{DB_{opt}L}{T} \tag{13}$$

where maximum delay time is largest delay between wavelengths arriving at the client and $T$ is the time between weights arriving the client, set by the system clock rate. For simulated values of $\chi = 0.05 - 0.1$ the temporal crosstalk factor does not significantly change the accuracy of classification on large MNIST models and Alexnet. Two different wavelength ranges are used for long distance communications: the optical C band (1550 nm) and optical O band (1310 nm). While we operate in the optical C band (1550 nm) for our demonstration, the O-band (1310 nm) offers near zero dispersion with comparable optical loss. Assuming standard single mode SMF-28 fiber [37], dispersion in the C-band is given by 18 ps/(nm*km) while in the O-band it is given by:

$$D = 0.092 \frac{ps}{(nm)^2 km} |\lambda - 1314nm| \tag{14}$$

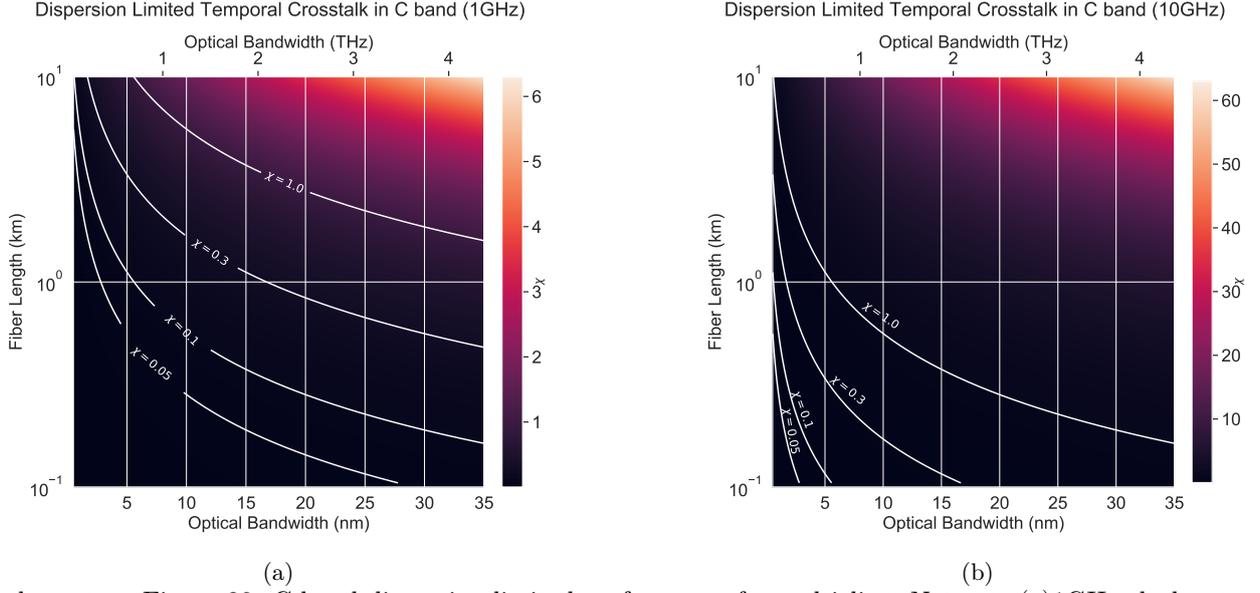

(a)

(b)

Supplementary Figure 20: C-band dispersion limited performance for multiclient Netcast. (a)1GHz clock rate. (b) 10GHz clock rate

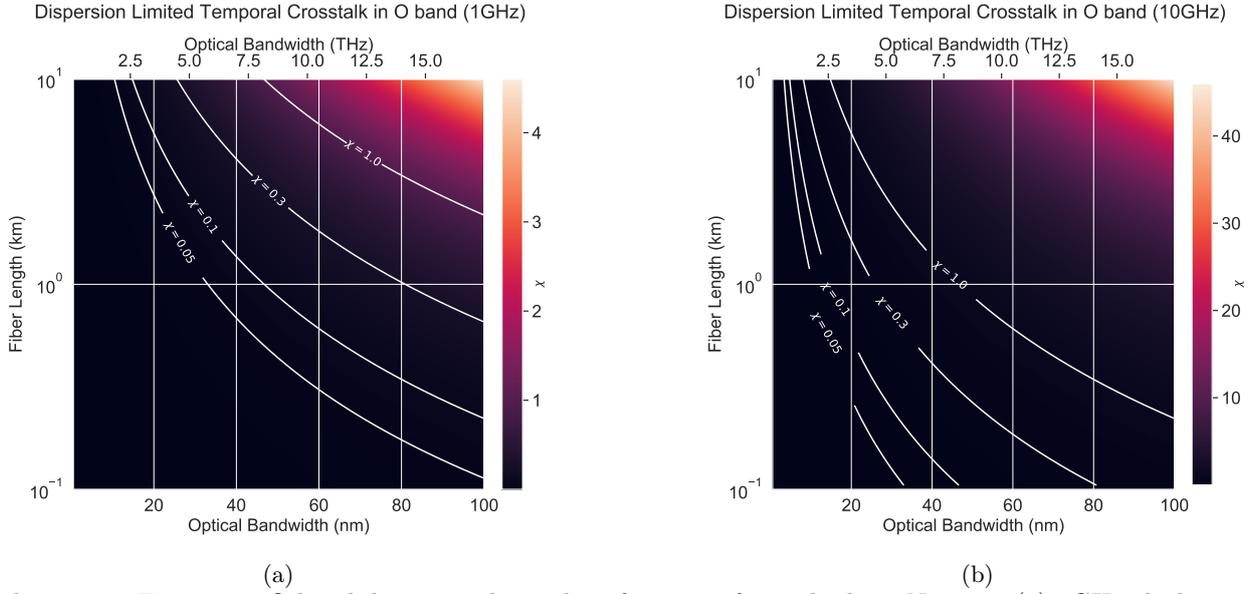

(a)

(b)

Supplementary Figure 21: O-band dispersion limited performance for multiclient Netcast. (a) 1GHz clock rate. (b) 10GHz clock rate

Using these equations we model the effects of temporal crosstalk as a function of optical bandwidth and fiber length at 1GHz and 10GHz in both the C-band (Supplementary Figure 20) and O-band (Supplementary Figure 21).

The above modeling shows that operation in the C-band (1550 nm) can lead to limited range and optical bandwidth when distributing to multiple clients. However, the O-band (1310 nm) can enable high clock rates (10 GHz) with THz of bandwidth on fiber links as long as 10km to multiple clients simultaneously.

Operation close to zero dispersion brings up reasonable questions about nonlinear effects in fiber, as zero dispersion means four-wave mixing effects can be phase matched over long distances. Experimental work has demonstrated DWDM communication systems in the zero-dispersion section of the O-band with low crosstalk from four wave mixing (0.3 dB) using 25 km of standard fiber and launch powers of $\approx$ 100 uW [38]. Operating at aJ/MAC of optical energy per wavelength at the receiver with 100 GHz clock speed would imply a 100 nW signal per wavelength at the client. Even accounting for attenuation of 20 dB over the fiber link this power is still an order of magnitude lower than the launch power reported in the above paper. To understand the effect this order of magnitude has on Netcast we note that four wave mixing scales as:

$$P_{FWM} \propto P_1(\omega_1)P_2(\omega_2)P_3(\omega_3) \tag{15}$$

where $P_{FWM}$ is the generated power at a chosen four wave mixing wavelength and $P_{1,2,3}$ are launch powers at frequencies $\omega_{1,2,3}$ respectively. Decreasing the launch power of each wavelength into the fiber by a factor of $\xi$ decreases the effect of four wave mixing by $\xi^3$. Similarly, any other third order nonlinear optical effect will be reduced by a factor of $\xi^3$. An order of magnitude reduction in launch power means $\xi = 10$ and implies $P_{FWM}$ is lowered by a factor of 1000, implying negligible nonlinear effects.

## XXI. REPRESENTATION OF THERMAL NOISE VS SHOT NOISE

In other literature different equations can be found which also represent thermal noise and shot noise. Here we discuss and derive these equations from the two equations used for thermal and shot noise in the main text to show readers that both representations are equal and allow readers to more easily back-of-the-envelope shot and thermal noise for their systems.

### A. Thermal Noise

Thermal noise scales as $\sigma_{th} = \frac{\sqrt{kTC}}{q}$ electrons of readout noise [39]. A common represetation for thermal noise is in units of RF power $P_{th} = 4kT\Delta f$ which we will now derive. $\frac{\sqrt{kTC}}{q}$ can be converted to units of readout voltage $V_{rms} = \sqrt{\frac{kT}{C}}$. Capacitors themselves do not have a thermal noise, the thermal readout noise of a capacitor is associated with the finite impedance of the conductors they are connected to. For an RC electrical circuit the noise bandwidth is given by $\Delta f = \frac{1}{4RC}$ [40]. Substituting into our original expression we find that the thermal noise in number of readout electrons in units of resistance and bandwidth is $\sigma_{th} = \frac{\sqrt{\frac{kT}{4R\Delta f}}}{q}$. We convert to units of power:

$$P_{th} = \frac{V_{rms}^2}{R} = \frac{kT}{RC} = 4kT\Delta f \tag{16}$$

This thermal noise power is independent of the resistance of the conductor connected to the capacitor. For an amplified photodetector that performs readout on each multiply the optical energy per MAC can be found with:

$$E_{mac} = \text{SNR} \cdot \eta \frac{h\nu}{q} Q_{\text{mac}} = \text{SNR} \cdot \eta \frac{h\nu}{q} \frac{I_{\text{rec}}}{\Delta f} = \text{SNR} \cdot \eta \frac{h\nu}{q} \frac{V_{\text{rec}}}{R\Delta f} = \text{SNR} \cdot \eta \frac{h\nu}{q} \sqrt{\frac{4kT}{R\Delta f}} \tag{17}$$

where $I_{rec}$ and $V_{rec}$ are the receiver photocurrents and generated voltage respectively. SNR is the required signal to noise ratio for accurate clasisifcation on a given neural network model with typical values between 10 and 100, as is discussed and shown in [30]. Assuming values for conventional telecom receivers $\Delta f \approx 10\text{GHz}$ and $R \approx 1000$ we find that the $E_{mac} \approx \text{SNR} \cdot 50\text{aJ}$ [41, 42]. Assuming a required signal to noise ratio of $\approx 20$ the fJ/MAC value from the main text is found.

While these equations involving resistance and bandwidth are useful for engineering purposes, they do fundamentally hide how we can design better receiver systems. As is noted at the top of this section, we can express thermal noise in terms of the receiver capacitance, with decreasing capacitance lowering the number of thermal noise electrons on readout. Capacitance decreases with the size of a semiconductor or metal [43]. As a result, to enable the best possible photoreceivers photonics and electronics should be kept as close as possible together, within a few micrometers. Emerging technologies from foundry platforms such as globalfoundries and IHP enable transistors within a few micrometers of silicon photonic components including modulators, waveguides, and photodetectors [44, 45].

### B. Shot Noise

Shot noise originates from the quantum nature of light. Mathematically, it is described by Poisson statistics:

$$P_{shot}(q) = \frac{e^{-n_p}(n_p^q)}{q!} \tag{18}$$

where $n_p$ is the mean photon number in a measurement window at the receiver and q number of photons we wish to know the probability of detecting. This distribution has the property that it's expected value and variance are equal, meaning that the signal to noise ratio of a shot-noise limited measurement can be described as $\sqrt{n_p}$. Using this information we can derive standard equations that describe shot noise in electronics textbooks. Assuming unity quantum efficiency the rms electron shot noise is $\sigma_{elec,rms} = \sqrt{2\frac{Q_{mac}}{q}}$ where $Q_{mac}$ is the average charge per MAC and the factor of 2 originates from the double sided nature of the noise spectral density. Converting to units of current we find $\sigma_{elec,rms} = \sqrt{\frac{2I}{q\Delta f}} = \frac{\sqrt{2qI\Delta f}}{q\Delta f}$. Converting to current we find $\sigma_i = \sqrt{2qI\Delta f}$ which is listed in most electronics textbooks [46].

Shot noise is a fundamental noise source and can not be overcome through time-integration. Poisson random variables have the property that two independent poisson random variables $X_1, X_2$ with mean value $\mu_1, \mu_2$ will add together to form $X_3$ with mean value $\mu_1 + \mu_2$. As a result, once a measurement from either a time-integrator or TIA is shot-noise limited it can not be improved further.

### C. Comparing Thermal Noise and Shot Noise

One consideration is in what regimes thermal noise dominates and in what regime shot noise dominates. To quantify this we take the ratio of the expressions for shot and thermal noise:

$$\sigma_{shot}/\sigma_{th} = \frac{\sqrt{n_{ph}}}{\frac{\sqrt{kTC}}{q}} \tag{19}$$

To find this crossover we assume $\sigma_{shot} = \sigma_{th}$. We find that the mean number of photons per readout to reach a shot-noise limited regime for a fixed receiver capacitance is $n_{ph} = \frac{kTC}{q^2}$. For $C = 10pF$ used in the main text this would be $n_{ph} = 1.6 * 10^6$, which would be a shot-noise limited SNR of $\approx 1000$, far larger than required for the application. Two pieces of information can be gained by this equation: 1. A measurement being shot-noise limited does not imply it is the best operating regime for neural network computation. Any photoreceiver can be made shot noise limited by applying sufficient optical power within a given integration window, but the signal to noise ratio at the receiver may be orders of magnitude larger than what is required for the application, implying laser power is being wasted or operating bandwidth can increase. 2. The shot-noise to thermal noise crossover point scales linearly with capacitance. Decreasing capacitance makes it easier to reach a shot-noise limited regime.

## D. Dark Current

Similar to the above, dark current of a detector presents another potential noise source. Dark current is caused by defects the photodiodes absorption material generating excess electro-hole hairs even when no optical illumination is incident. Dark current electrons and holes are thermally excited, so this current source looks similar to a thermal source for the sake of modelling. The photodiode we use, FGA01FC, has a quoted dark current of 50pA. When measuring the optical energy per MAC of the time-integrating receiver we operate sufficiently fast (>10MHz) that the signal we are integrating dominates over the dark current, which we do not observe.

## E. Flicker ($\frac{1}{f}$) Noise

Photonic and electronic components exhibit a low frequency noise, flicker noise or $\frac{1}{f}$ noise, that in certain applications can dominate shot-noise. This noise source originates from material properties such as defects and is fabrication dependent [46]. There are several models of flicker noise, but for our discussion we will assume that flicker noise follows a spectral distribution of:

$$S_I(\frac{A^2}{Hz}) = K_f I^\alpha / f^\beta \tag{20}$$

where $\alpha \approx 2$ and $\beta \approx 1$ in most situations [46, 47], $K_f$ is a unitless material/device dependent parameter and $I$ is the current flowing through the device under consideration. In modern CMOS and BiCMOS processes $K_f$ is in the range $\approx [10^{-7}, 10^{-9}]$ [47, 48].

At higher frequencies flicker noise decreases below the shot-noise floor. This crossover point is characterized by a frequency $f_c$ given by:

$$\frac{K_f I^2}{f_c} = 2qI \implies f_c = \frac{K_f I}{2q} \tag{21}$$

from this crossover point we observe that systems with higher optical and electrical powers, such as standard transceivers, will experience more flicker noise. However, for the low-power electronics proposed in the main text for next generation time-integrating receivers the flick noise corner will be orders of magnitude lower since currents, both optical and electrical , will be $I \approx 1uA$. Deployed systems will operate a high-speed, away from the low-frequency $\frac{1}{f}$. As a reference, the reset time of the integrator used in the main text is 10us, corresponding to a low frequency cutoff of $\approx 100kHz$ readout bandwidth. For a 1uA optical of electrical current with $K_f = 10^{-8}$ the flicker-noise to shot noise crossover is at $f_c \approx 30kHz$.

$K_f$ may also further improve the future through advances in materials processing.

## XXII. NETCAST AT RADIO FREQUENCIES

In this text we have proposed Netcast using fiber and free-space optics. However, the most ubiquitously deployed device in our daily lives, the cell phone, does not have a native optical link. While one could in principle integrated custom optical receivers into cellphones to enable Netcast is may be more practical to make use of existing 5G infrastructure to deploy weight data to radio frequencies (RF) devices. To see if this idea is feasible we perform the following calculation: The power at an RF device before amplification is $P_{rec} = \sqrt{P_s P_{LO}}$ where $P_s$ is the power of the received RF signal (typically pW scale for 5G systems [49]) and $P_{LO}$ is the power of the local oscillator. Assuming a deployment bandwidth of 1GHz and a local oscillator power of $1\mu$W we can find the RF power SNR at the receiver as:

$$\text{SNR} = \frac{\text{P}_{\text{rec}}}{\text{P}_{\text{noise}}} = \frac{\text{V}_{\text{rec}}^2}{\text{V}_{\text{noise}}^2} = \frac{\sqrt{P_s P_{LO}}}{\text{kT}\Delta\text{f}} \tag{22}$$

where the denominator of this equation is thermal noise. This evaluates to a power SNR of 241 or a voltage SNR of 15, sufficient for our applications. Both bandwidth and receiver sensitivity can be improved by time-integrating the homodyne signal. Assuming we only see thermal readout noise once after $M$ timesteps then a similar analysis would imply that 1pW of received power RF power with a $1\mu$ W LO could enable 100GHz of accurate computation, assuming there is sufficient spectrum at the carrier frequency of interest. Another benefit of RF communication architecture is the large constellation diagrams that are in use. Existing 5G systems utilize 1024 QAM, allowing for a large number of amplitude and phase controlled levels [49].

## XXIII. DEPLOYMENT OF NETCAST TO SPACECRAFT

An interesting application of Netcast is in deploying neural network models to satellites for image classification. The challenge associated with deploying a model to a satellite is the loss from free-space propagation associated with diffraction of the optical signal. This effect is modeled by the Friis transmission equation [50]:

$$P_r = P_t \frac{A_t A_r}{\lambda^2 R^2} \tag{23}$$

where $P_r$ is the optical power received at the client, $P_t$ is the optical power transmitted from a base station, $A_t$ is the transmitted effective aperture area, $A_r$ is the receiver effective aperture area, $\lambda$ is the wavelength of light used, and $R$ is the distance communicated.

We estimate the order of magnitude of performance Netcast can offer us by considering two scenarios: deployment to a low-earth-orbit satellite and deployment to a satellite located around Mars.

Low earth orbit around earth is located $R \approx 2,000$ km above the earth's surface. Satellites deployed in low earth orbit could be running neural network models associated with image classification of weather, hyperspectral imaging data, or disaster relief. We assume we transmit at standard telecommunication wavelengths ($\lambda = 1550nm$) with $P_t = 1$ watt of power on each of N = 100 wavelengths. We assume effective transmitted and receiver aperture areas of $A_t = A_r = 0.1\text{m}^2$. For each wavelength, the power received by the satellite would be $\approx$ 1mW. Assuming operation of $10^{-19}$J/MAC this amount of optical power could enable $10^{16}$ MAC/s per wavelength on a low-earth-orbit satellite, for a theoretical maximum bandwidth of $10^{19}$MAC/s using all wavelengths. Practical engineering constraints, such as finite modulator bandwidth

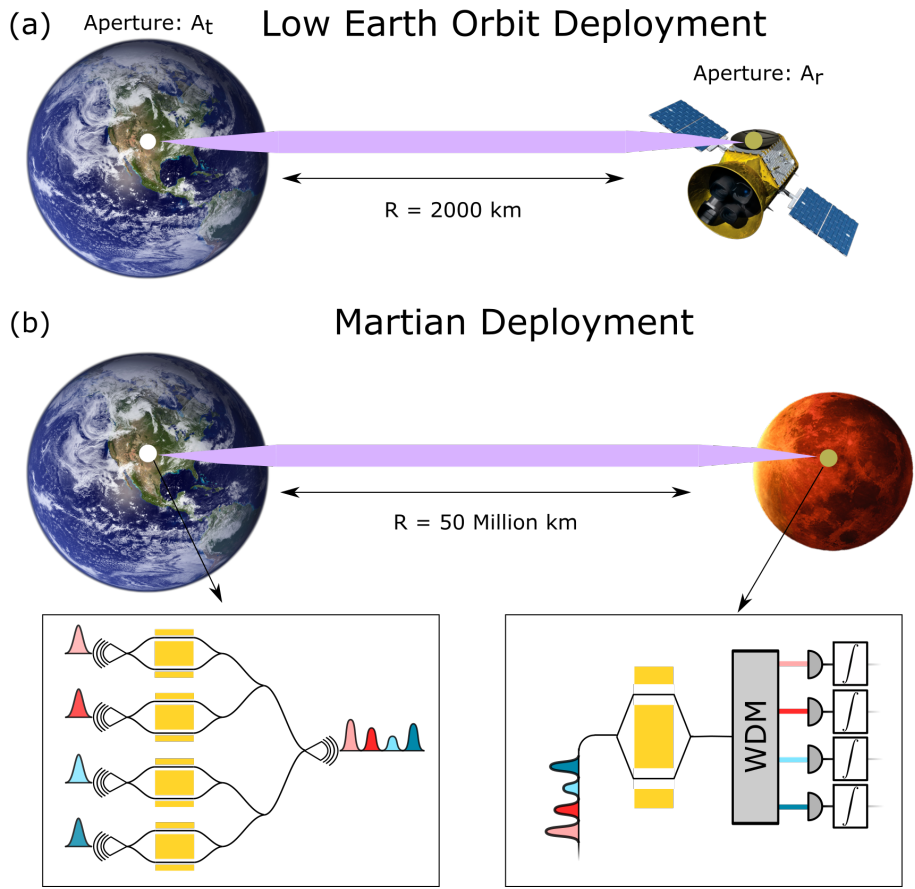

Supplementary Figure 22: Example of space deployment of Netcast. (a) Terrestrial basestation containing a weight server transmits weight data to a satellite located in low earth orbit. While transmitting from the base station to satellite the optical beam diffracts. (b) In a longer distance deployment, such as to mars, the beam significantly diffracts through propagation, leading to significant loss

may pose near term scaling challenges to achieving this theoretical throughput, but this high possible throughput gives a roadmap for improvement.

Satellites deployed far away from the earth, such as those located at Mars, are used for basic science such as searching for water and signs of life on distant planets. Mars is located $\approx 50 \cdot 10^9$ meters from earth. Assuming the same specifications for transmitted and receiver as above we find that the received power at the martian satellite will be $\approx 10^{-12}$ W per wavelength. This enables $\approx 10^9$ MAC/s at the receiver assuming shot noise limited detectors and $N = 100$ wavelengths. While initially disappointing, this result can be improved through significant engineering of the transmitted and receiver apertures, using more transmission wavelengths, higher transmitter powers, or moving to shorter wavelengths. For example, consider a future base station capable of generating and modulating many wavelengths ($N = 1000$) centered around 532nm (green) at GHz of speed with 10 watts of power per wavelength. Assuming the same transmit and receive aperture size $\approx 140$pW of light will be received per wavelength, enabling 1.4TeraMAC/s of compute power using all wavelengths. Other future technological developments, such as transmitters which beat the diffraction limit utilizing superdirectivity [51], may improve this computing result by orders of magnitude.

## XXIV. ENCODING NEGATIVE NUMBERS

In the main text we use a system that only performs computation on non-negative floating point values. Optical intensity does not have a negative form, so another technique is required to encode floating point values. Recent work with a similar problem has proposed shifting floating point zero to the middle of the transmission range of the modulator [31]. I'll briefly describe this shifting method here: We wish to encode input and weight data in the range $[x_{min}, x_{max}]$, $[w_{min}, w_{max}]$ where $x_{min}, w_{min} < 0$ and $x_{max}, w_{max} > 0$. Because we can only encode non-negative values into our system we linearly shift the floating point values up by the minimum of their range, resulting in values $x + x_{min}, w + w_{min}$ in the ranges $[0, x_{min} + x_{max}]$ and $[0, w_{min} + w_{max}]$ respectively. Multiplying these two values we obtain $y_{shifted} = y + w \cdot x_{min} + x \cdot w_{min} + x_{min} \cdot w_{min}$ where y is the floating point product $x \cdot w$. Then, in post, the other terms are calculated and subtracted from the measured value, leaving just $y$. If a form of accumulation is used, such as time integration in this paper or spatial integration from the cited paper, then the measured value is $\sum_i x_i w_i + x_{min} \sum_i w_i + w_{min} \sum_i x_i + \sum_i x_{min} w_{min}$ where the first term is the output activation. There are practical ways to obtain the remaining terms such as tapping off the weight values at the client and time integrating their signal, so the computational overhead of implementing this method on hardware can be small. However, an issue with encoding floating point zero as anything other than the absence of photons comes from the systems resilience to systematic errors or calibration errors. Suppose we substitute a small calibration error $\epsilon$ into both the input and weight modulator as shown in Supplementary Figure 23. Without this shifting method error scales as:

$$\sum_i (x_i + \epsilon)(w_i + \epsilon) = \sum_i x_i w_i + \epsilon x_i + \epsilon w_i + \epsilon^2 \qquad (24)$$

where the $\epsilon^2$ can be discarded for being small. For the shifting technique used in the above reference error scales as:

$$\sum_i (x_i + \epsilon)(w_i + \epsilon) = \sum_i x_i w_i + \epsilon x_i + x_i w_{min} + \epsilon w_i + \epsilon^2 + \epsilon w_{min} + x_{min} w_i + \epsilon x_{min} + x_{min} w_{min} \qquad (25)$$

Where the terms $\sum_i \epsilon w_{min}$ and $\sum_i \epsilon x_{min}$ are of interest. These terms, when compared with $\sum_i \epsilon x_i$ and $\sum_i \epsilon w_i$ are much larger because of neural network's inherent sparsity, which can reach beyond 90% [52]. This means these introduced terms will be an order of magnitude larger than the prior error terms. Conceptually, this can be understood as the zero float point of both modulators shifting by epsilon, leading to a build up in imprecision at the receiver. Because the origin of this is in the sparsity of neural networks, we call this problem the "fat zero" problem because of the large number of zero values in neural network compute.

Netcast has many ways to encode non-negative values in the intensity of light, with some examples shown in Supplemental Figure 23(b). For example, one can make use of either distinct wavelengths where one wavelength encodes positive values and the other negative values, which would be reliable, but at the cost of decreased optical bandwidth and extra hardware at the weight server. Another proposal would be to make use of orthogonal polarizations in the optical medium, such as TE and TM polarization for positive and negative values respectively. One could also make use of distinct spatial modes, if allowed by the target application, where either two orthogonal modes in the medium or two cores in a multicore fiber encode positive and negative values respectively. In these proposed solutions, an optical attenuator must be added to reduce the intensity of light when trying to encode low floating point values such as zero. We do this for conceptually the same reason the fat-zero problem occurs: any imperfection in the splitting ratio of the modulator would manifest itself as a zero-error at the output, leading to error.

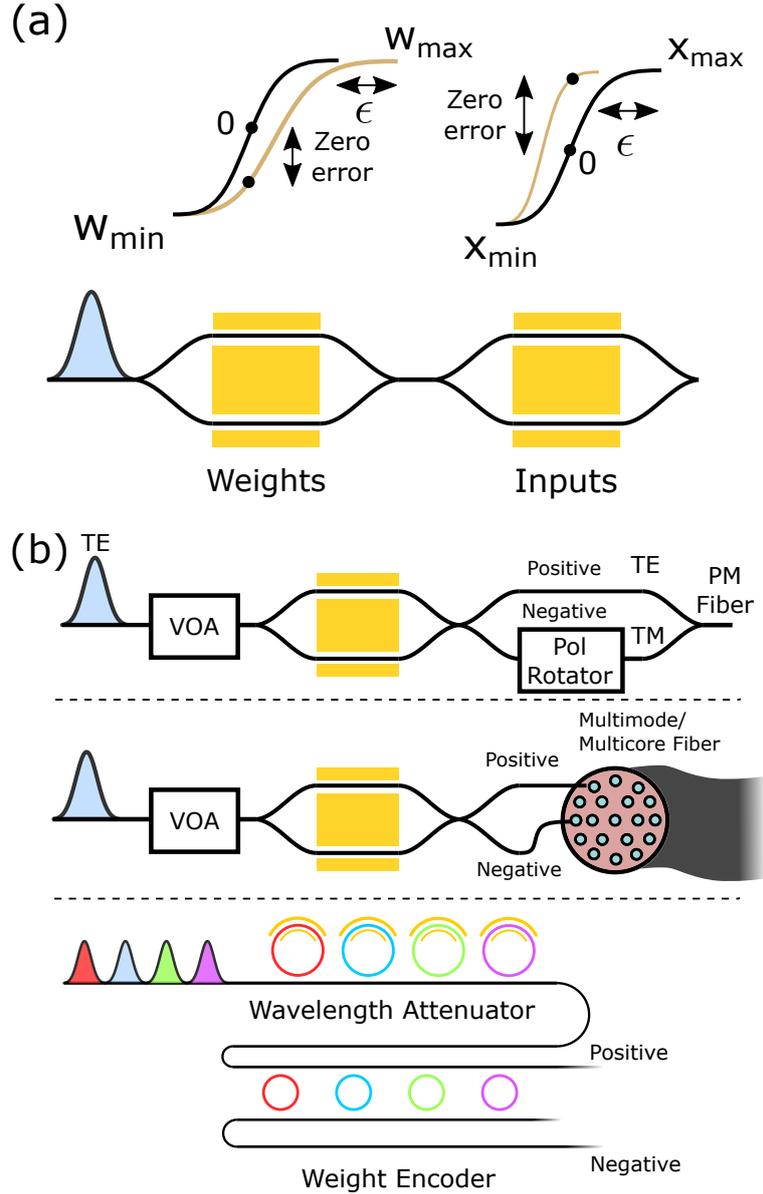

Supplementary Figure 23: Encoding Negative Numbers with Netcast (a) An input and weight modulator encode floating point values between $[x_{min}, x_{max}]$ and $[w_{min}, w_{max}]$ respectively. When the zero floating point is encoded in the middle of the transfer function it becomes sensitive to any drift in calibration, $\epsilon$. (b) Methods for encoding negative values in Netcast. Polarization diversity, spatial encoding, or wavelength multiplexing could be used to encode positive and negative weight values. Attenuators, such as variable optical attenuators (VOAs) are used to make sure zero is encoed as the absence of photons.

## XXV. POLARIZATION DRIFT AND MODE DISPERSION

Most fiber used for practical applications is non-polarization maintaining. As a result, propagating through the fiber will scramble the polarization. This leads to issues if, for example, the client is sensitive to polarization. Further, fibers have a polarization mode dispersion, causing both TE and TM polarizations to become temporally delayed from each other [37]. This polarization mode dispersion value is small, typically $0.1~\mathrm{ps}/\sqrt{km}$. Even for large fiber links (100 km) this does not lead to enough temporal crosstalk to change model accuracy. However, this time delay does lead to a significant phase difference between the TE and TM modes (tens to hundreds of radians). This phase could vary from wavelength to wavelength, requiring $N = 100$ active components to phase stabilize the two polarizations at the client. Here, we propose a method which is purely passive and broadband, allowing for the client to be resistant to polarization drift and mode dispersion simultaneously. Supplementary Figure. 24 shows the key idea, where each wavelength is composed of a random superposition of TE and TM modes. At the client they pass through a broadband polarization splitter and rotator, passively splitting TE and TM into separate waveguides and rotating TM to TE. Then, these two TE modes enter opposite ends of a broadband modulator (MZM) where they both receive the same attenuation before exiting on opposite ends. Each result is detected on photodetectors and the generated photocurrents are summed together by tying the photodetectors electrodes together. To avoid spurious backreflections a non-reciprocal device, such as an isolator or circulator, must be added to the client.

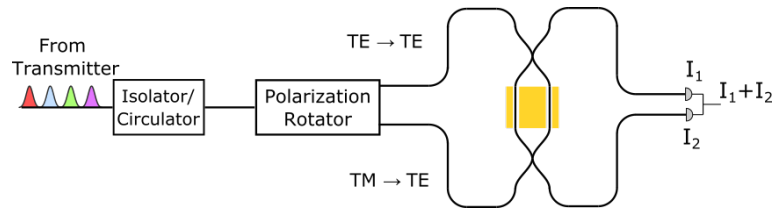

Supplementary Figure 24: Overcoming polarization mode dispersion and polarization drift at the client

We note here that this scheme will not work if the client modulator makes use of traveling wave electrodes as the RF wave and one of the optical waves will propagate in different directions. Traveling wave electrodes are not needed for high speed operation of the client side modulator assuming the modulator is sufficiently short and drive electronics are co-integrated.

Another method for implementing a similar fix is to make use of two copies of the receiver, one for TE and another for TM polarization. This doubles receiver hardware and energy consumption, but gives more tolerance to things like on-chip backreflection from imperfect optical components.

## XXVI. COHERENT DETECTION WITHOUT PHASE STABILIZATION

Coherent detection gives the client a significant gain over receiver thermal noise, enabling lower numbers of receiver photons per MAC. However, coherent detection also comes with a cost, the phase of the client-side local oscillator and receiver signal must stay at a fixed phase difference. In practice, deployed fibers are not stable and the phase over deployed fiber drifts significantly. Modern coherent transceivers overcome this problem by making use of complex digital signal processing (DSP) electronics which can compensate for or track this phase drift in time. These DSP ASICs consume watts of power, making them insufficient for client side energy requirements. In addition to phase drift, the receiver must lock the frequency of it's local oscillator laser to the transmitted optical signal. While techniques such as injection locking could be used to lock the transmitter and receiver lasers, they take away received light, lowering the receivers effective optical sensitivity. Here, we propose a method for overcoming shifts in phase by making use of sum-squared detection. Shown in Supplementary Figure 25 is the proposed scheme, where sum-squared detection is used to take the difference and phase and frequency and ignore them by using an IQ receiver (complex hybrid/90 degree hybrid). Each output of the IQ receiver is then detected and the I and Q component have sum-squared detection performed, removing the phase and frequency componentso of the detection.

A downside of this approach is that we can no longer make use of phase as a degree of freedom to encode information in. However, the complexity of stabilize frequency and phase between the transmitter and client is reduced by such a large margin that this tradeoff is preferable. Another downside of this approach is that it requires high speed detectors

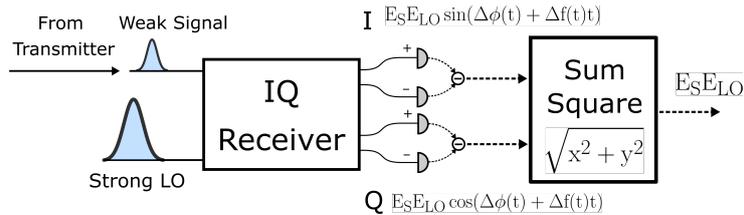

Supplementary Figure 25: Coherent detection without the need for phase stability.

to be able to see $\Delta f$, since $\Delta f$ will likely be GHz for free-running tunable lasers.

## XXVII. ALTERNATIVE ARCHITECTURES FOR NETCAST

Netcast is fundamentally an insight in using time-frequency encoding and other techniques to lower client energy consumption. We already mentioned in Supplementary Materials XII that coherent detection can be incorporated at the client to give significant gains to receiver sensitivity. The hardware and architecture utilized in the main text is simply one method of performing Netcast. Supplementary Figure 26 showcases other Netcast architectures. Supplementary Figure 26(c) shows an alternative to Netcast that makes use of frequency-integrating receivers rather than time-integrating receivers. A frequency-integrating receiver is a high-speed amplified photodetector with a broad absorption bandwidth. In this frequency-integration time-separation scheme (FITS) columns of the weight matrix are cast to separate optical wavelengths rather than time. In this scheme only a single (or two if positive and negative values are encoded simultaneously) photodetector needs to be used. At the receiver a bank of low-speed tunable wavelength filters (such as cavity modulators) encode input activation values. The rows of the weight matrix are deployed on each timestep and readout on each timestep. After $N$ timesteps the $Y_N$ output activation vector is generated at the receiver. This FITS scheme offers similar performance on many metrics when compared to TIFS, but some assumptions must be made. Devices such as DACs, ADCs, and modulators are all amortized by factors of $N$ or $M$ respectively in this scheme, with the results listed in Supplementary Table I. In both the TIFS and FITS schemes shown in Supplementary Figure 26(b,c) resonant filters are used to encode input activation data. One concern with using resonant devices is that after fabrication they have a resonance variation typically on the order of $\approx$1nm, larger than the electro-optic tuning range of standard PDK cavity modulators [53]. To compensate this, several technologies can be employed, such as implanting germanium into the cavity and thermally annealing [54], laser trimming of cavities by growing thermal oxide [55], or advances in the phase tuning range of depletion mode cavities through either better doping design or higher finesse [56–58].

| Netcast FITS Client Energy Consumption | | | | |
|---|---|---|---|---|
| Device | Number of Devices | Fan-out | Energy per Device | Energy per MAC |
| Modulator [56] | M | $N$ | $\sim 1$ fJ | $\sim$ M fJ$/N$ |
| DAC [59] | 1 | $N$ | $\sim 1$ pJ | $\sim 1$ pJ$/N$ |
| ADC [60] | 1 | $M$ | $\sim 1$ pJ | $\sim 1$ pJ$/M$ |
| Laser | 1 | – | – | $\sim 1$ aJ |
| Total | – | – | – | $\sim 1$ pJ$/N$ |

Table I: Device contributions to FITS client performance assuming conventional technology.

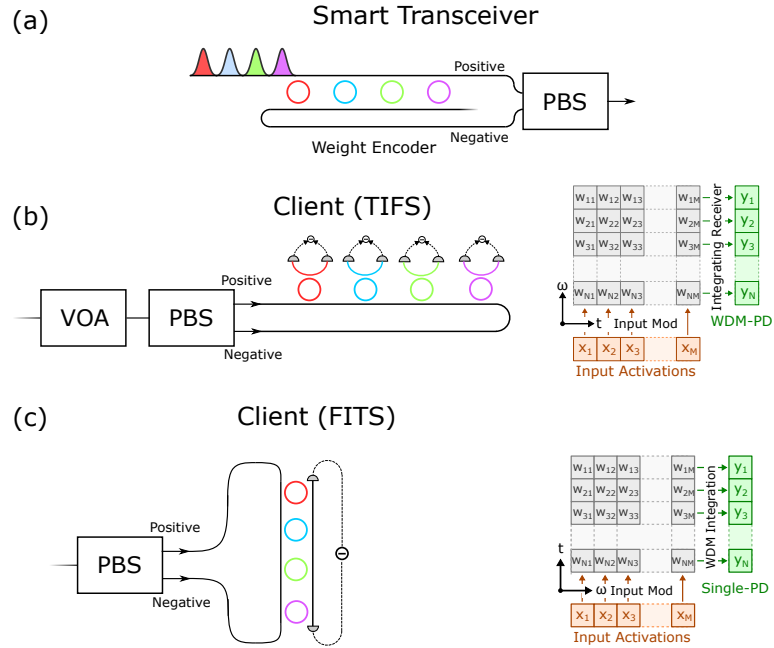

Supplementary Figure 26: Other implementations of Netcast. (a) A smart transceiver composed of resonant add-drop filters which allow for positive and negative weights to be encoded onto a polarization maintaing (PM) fiber.(b) Time-integration frequency-separation (TIFS) netcast using resonant devices, similar to the architecture in the main text. (c) Frequency-integration time-separation (FITS) netcast where frequencies are summed in time.

---

[1] R. Ding, Y. Liu, Q. Li, Y. Yang, Y. Ma, K. Padmaraju, A. E.-J. Lim, G.-Q. Lo, K. Bergman, T. Baehr-Jones, *et al.*, Optics communications **321**, 124 (2014).

[2] N. M. Fahrenkopf, C. McDonough, G. L. Leake, Z. Su, E. Timurdogan, and D. D. Coolbaugh, IEEE Journal of Selected Topics in Quantum Electronics **25**, 1 (2019).

[3] J. Komma, C. Schwarz, G. Hofmann, D. Heinert, and R. Nawrodt, Applied Physics Letters **101**, 041905 (2012).

[4] N. C. Harris, Y. Ma, J. Mower, T. Baehr-Jones, D. Englund, M. Hochberg, and C. Galland, Optics express **22**, 10487 (2014).

[5] R. Baghdadi, M. Gould, S. Gupta, M. Tymchenko, D. Bunandar, C. Ramey, and N. C. Harris, Optics Express **29**, 19113 (2021).

[6] K. Giewont, K. Nummy, F. A. Anderson, J. Ayala, T. Barwicz, Y. Bian, K. K. Dezfulian, D. M. Gill, T. Houghton, S. Hu, *et al.*, IEEE Journal of Selected Topics in Quantum Electronics **25**, 1 (2019).

[7] L. Chen, D. Schwarzer, J. A. Lau, V. B. Verma, M. J. Stevens, F. Marsili, R. P. Mirin, S. W. Nam, and A. M. Wodtke, Optics Express **26**, 14859 (2018).

[8] D. Bunandar, A. Lentine, C. Lee, H. Cai, C. M. Long, N. Boynton, N. Martinez, C. DeRose, C. Chen, M. Grein, *et al.*, Physical Review X **8**, 021009 (2018).

[9] M. Prabhu, C. Errando-Herranz, L. De Santis, I. Christen, C. Chen, and D. R. Englund, arXiv preprint arXiv:2202.02342 (2022).

[10] B. Razavi, IEEE Solid-State Circuits Magazine **9**, 9 (2017).

[11] H. Khorramabadi, Uc berkley ee237 lecture note 10 (2008).

[12] G. Mourgias-Alexandris, A. Tsakyridis, N. Passalis, and M. Kirtas, in *European Conference on Optical Communications*, IKEECONF-2021-404 (Aristotle University of Thessaloniki, 2021).

[13] M. J. Byrd, E. Timurdogan, Z. Su, C. V. Poulton, N. M. Fahrenkopf, G. Leake, D. D. Coolbaugh, and M. R. Watts, Optics letters **42**, 851 (2017).

[14] S. Lischke, A. Peczek, J. Morgan, K. Sun, D. Steckler, Y. Yamamoto, F. Korndörfer, C. Mai, S. Marschmeyer, M. Fraschke, *et al.*, Nature Photonics **15**, 925 (2021).

[15] M. Streshinsky, A. Novack, R. Ding, Y. Liu, A. E.-J. Lim, P. G.-Q. Lo, T. Baehr-Jones, and M. Hochberg, Journal of Lightwave Technology **32**, 4370 (2014).

[16] B. B. Bakir, A. V. de Gyves, R. Orobtchouk, P. Lyan, C. Porzier, A. Roman, and J.-M. Fedeli, IEEE Photonics Technology Letters **22**, 739 (2010).

[17] J. Notaros, F. Pavanello, M. T. Wade, C. M. Gentry, A. Atabaki, L. Alloatti, R. J. Ram, and M. A. Popović, in *2016 Optical Fiber Communications Conference and Exhibition (OFC)* (IEEE, 2016) pp. 1–3.

[18] Y. Ding, C. Peucheret, H. Ou, and K. Yvind, Optics letters **39**, 5348 (2014).

[19] S. Bandyopadhyay and D. Englund, arXiv preprint arXiv:2110.12851 (2021).

[20] S. Yu, L. Ranno, Q. Du, S. Serna, C. McDonough, N. Fahrenkopf, T. Gu, and J. Hu, arXiv preprint arXiv:2112.14357 (2021).

[21] A. Yariv and P. Yeh, *Photonics: optical electronics in modern communications* (Oxford university press, 2007).

[22] R. Hamerly, L. Bernstein, A. Sludds, M. Soljačić, and D. Englund, Physical Review X **9**, 021032 (2019).

[23] L. Lundberg, M. Mazur, A. Mirani, B. Foo, J. Schröder, V. Torres-Company, M. Karlsson, and P. A. Andrekson, Nature communications **11**, 1 (2020).

[24] J. Schröder, A. Fülöp, M. Mazur, L. Lundberg, Ó. B. Helgason, M. Karlsson, P. A. Andrekson, *et al.*, Journal of Lightwave Technology **37**, 1663 (2019).

[25] Y. Geng, X. Han, G. Deng, Q. Zhou, K. Qiu, and H. Zhou, in *2021 Optical Fiber Communications Conference and Exhibition (OFC)* (IEEE, 2021) pp. 1–3.

[26] Y. Hu, M. Yu, B. Buscaino, N. Sinclair, D. Zhu, R. Cheng, A. Shams-Ansari, L. Shao, M. Zhang, J. M. Kahn, *et al.*, arXiv preprint arXiv:2111.14743 (2021).

[27] A. L. Gaeta, M. Lipson, and T. J. Kippenberg, nature photonics **13**, 158 (2019).

[28] L. Chang, S. Liu, and J. E. Bowers, Nature Photonics **16**, 95 (2022).

[29] A. Yamamoto, T. Okaniwa, Y. Yafuso, and M. Nishita, Furukawa Electric Review , 3 (2015).

[30] S. Garg, J. Lou, A. Jain, and M. Nahmias, arXiv preprint arXiv:2102.06365 (2021).

[31] T. Wang, S.-Y. Ma, L. G. Wright, T. Onodera, B. C. Richard, and P. L. McMahon, Nature Communications **13**, 1 (2022).

[32] C. R. Harris, K. J. Millman, S. J. Van Der Walt, R. Gommers, P. Virtanen, D. Cournapeau, E. Wieser, J. Taylor, S. Berg, N. J. Smith, *et al.*, Nature **585**, 357 (2020).

[33] E. Timurdogan, C. V. Poulton, M. Byrd, and M. Watts, Nature Photonics **11**, 200 (2017).

[34] K.-S. Hyun and C.-Y. Park, Journal of applied physics **81**, 974 (1997).

[35] R. Hamerly, A. Sludds, L. Bernstein, M. Prabhu, C. Roques-Carmes, J. Carolan, Y. Yamamoto, M. Soljačić, and D. Englund, in *2019 IEEE International Electron Devices Meeting (IEDM)* (IEEE, 2019) pp. 22–8.

[36] R. Hamerly, A. Sludds, S. Bandyopadhyay, L. Bernstein, Z. Chen, M. Ghobadi, and D. Englund, in *Emerging Topics in Artificial Intelligence (ETAI) 2021*, Vol. 11804 (International Society for Optics and Photonics, 2021) p. 118041R.

[37] Corning® smf-28® ultra optical fiber.

[38] K. Markowski, Ł. Chorchos, and J. P. Turkiewicz, Applied optics **55**, 3051 (2016).

[39] J. Pierce, Proceedings of the IRE **44**, 601 (1956).

[40] K. H. Lundberg, Unpublished paper **3**, 28 (2002).

[41] Selection table for optical transimpedance amplifiers: Parametric search: Analog devices.

[42] M. G. Ahmed, T. N. Huynh, C. Williams, Y. Wang, R. Shringarpure, R. Yousefi, J. Roman, N. Ophir, and A. Rylyakov, in *Optical Fiber Communication Conference* (Optical Society of America, 2018) pp. M2D–1.

[43] D. A. Miller, Journal of Lightwave Technology **35**, 346 (2017).

[44] M. Rakowski, C. Meagher, K. Nummy, A. Aboketaf, J. Ayala, Y. Bian, B. Harris, K. Mclean, K. McStay, A. Sahin, *et al.*, in *Optical Fiber Communication Conference* (Optical Society of America, 2020) pp. T3H–3.

[45] L. Zimmermann, D. Knoll, M. Kroh, S. Lischke, D. Petousi, G. Winzer, and Y. Yamamoto, in *Optical Fiber Communication Conference* (Optical Society of America, 2015) pp. Th4E–5.

[46] P. R. Gray, P. J. Hurst, S. H. Lewis, and R. G. Meyer, *Analysis and design of analog integrated circuits* (John Wiley & Sons, 2009).

[47] M. Seif, F. Pascal, B. Sagnes, J. Elbeyrouthy, A. Hoffmann, S. Haendler, P. Chevalier, and D. Gloria, in *2017 29th International Conference on Microelectronics (ICM)* (IEEE, 2017) pp. 1–4.

[48] M. Von Haartman, *Low-frequency noise characterization, evaluation and modeling of advanced Si-and SiGe-based CMOS transistors*, Ph.D. thesis, KTH (2006).

[49] will 5g replace wi-fi.

[50] H. T. Friis, Proceedings of the IRE **34**, 254 (1946).

[51] R. W. Ziolkowski, Physical Review X **7**, 031017 (2017).

[52] V. Sze, Y.-H. Chen, T.-J. Yang, and J. S. Emer, Proceedings of the IEEE **105**, 2295 (2017).

[53] A. Rizzo, A. Novick, V. Gopal, B. Y. Kim, X. Ji, S. Daudlin, Y. Okawachi, Q. Cheng, M. Lipson, A. L. Gaeta, *et al.*, arXiv preprint arXiv:2109.10297 (2021).

[54] H. Jayatilleka, H. Frish, R. Kumar, J. Heck, C. Ma, M. N. Sakib, D. Huang, and H. Rong, Journal of Lightwave Technology **39**, 5083 (2021).

[55] C. J. Chen, J. Zheng, T. Gu, J. F. McMillan, M. Yu, G.-Q. Lo, D.-L. Kwong, and C. W. Wong, Optics express **19**, 12480 (2011).

[56] E. Timurdogan, C. M. Sorace-Agaskar, J. Sun, E. Shah Hosseini, A. Biberman, and M. R. Watts, Nature communications **5**, 1 (2014).

[57] H. Gevorgyan, A. Khilo, M. T. Wade, V. M. Stojanović, and M. A. Popović, Photonics Research **10**, A1 (2022).

[58] K. Al Qubaisi, D. Onural, H. Gevorgyan, and M. A. Popović, in *2021 Optical Fiber Communications Conference and Exhibition (OFC)* (IEEE, 2021) pp. 1–3.

[59] B. M. Pietro Caragiulo, Clayton Daigle, Dac performance survey 1996-2020 (2022).

[60] B. Murmann, Adc performance survey 1997-2021 (2022).